\documentclass[useAMS,usenatbib]{mn2e}
\usepackage{graphicx}
\usepackage{epsfig} 
\usepackage{amssymb}
\usepackage{textcomp}

\def\H2{H\,{\sevensize{II}}}

\title[X-rays \& the 21-cm power spectrum]{Modification of the 21-cm power spectrum by X-rays during the epoch of reionization}
\author[Warszawski, Geil \& Wyithe]{L. Warszawski\thanks{lilaw@ph.unimelb.edu.au (LW)}, P.M. Geil\thanks{pgeil@ph.unimelb.edu.au (PMG)} and J.S.B. Wyithe\thanks{swyithe@unimelb.edu.au (JSBW)}\\
School of Physics, University of Melbourne, Parkville, VIC 3010, Australia\\}

\begin{document}

\date{\today}

\pagerange{\pageref{firstpage}--\pageref{lastpage}} \pubyear{2008}

\maketitle

\label{firstpage}

\begin{abstract} 

We incorporate a contribution to reionization from X-rays within analytic and semi-numerical simulations of the 21-cm signal arising from neutral hydrogen during the epoch of reionization. The relatively long X-ray mean free path (MFP) means that ionizations due to X-rays are not subject to the same density bias as UV ionizations, resulting in a substantive modification to the statistics of the 21-cm signal. We explore the impact that X-ray ionizations have on the power spectrum (PS) of 21-cm fluctuations by varying both the average X-ray MFP and the fractional contribution of X-rays to reionization.  In general, prior to the epoch when the intergalactic medium is dominated by ionized regions (H\,{\sevensize II} regions), X-ray-induced ionization enhances fluctuations on spatial scales smaller than the X-ray MFP, provided that X-ray heating does not strongly supress galaxy formation.  Conversely, at later times when \H2 regions dominate, small-scale fluctuations in the 21-cm signal are suppressed by X-ray ionization. Our modelling also shows that the modification of the 21-cm signal due to the presence of X-rays is sensitive to the relative scales of the X-ray MFP, and the characteristic size of \H2 regions.  We therefore find that X-rays imprint an epoch and scale-dependent signature on the 21-cm PS, whose prominence depends on fractional X-ray contribution.  The degree of X-ray heating of the IGM also determines the extent to which these features can be discerned. We further show that the presence of X-rays smoothes out the shoulder-like signature of \H2 regions in the 21-cm PS. For example, a 10\% contribution to reionization from X-rays translates to a 20--30\% modulation in the 21-cm PS across the scale of \H2 regions. We show that the MWA will have sufficient sensitivity to detect this modification of the PS, so long as the X-ray photon MFP falls within the range of scales over which the array is most sensitive ($\sim0.1$\,Mpc$^{-1}$). In cases in which this MFP takes a much smaller value, an array with larger collecting area would be required. As a result, an X-ray contribution to reionization has the potential to substantially complicate analysis of the 21-cm PS. On the other hand, a combination of precision measurements and modelling of the 21-cm PS promises to provide an avenue for investigating the role and contribution of X-rays during reionization.

\end{abstract}
 

\section{Introduction}
\label{sec:intro}
The process of hydrogen reionization started with ionized (\H2) regions around the first galaxies, which later grew to surround groups of galaxies. Reionization completed once these \H2 regions overlapped and occupied most of the volume between galaxies. The inhomogeneous nature of the reionization process suggests that the spatial dependence of the statistics of redshifted 21-cm fluctuations will provide the most powerful probe of the reionization epoch \citep{FOB06}. With this as motivation, several experiments are currently under development that aim to detect the 21-cm signal during reionization, including the Low Frequency Array\footnote{http://www.lofar.org/} (LOFAR), the Murchison Widefield Array\footnote{http://www.haystack.mit.edu/ast/arrays/mwa/} (MWA) and the Precision Array to Probe Epoch of Reionization\footnote{http://astro.berkeley.edu/$\sim$dbacker/eor/} (PAPER), and more ambitious designs are being planned such as the Square Kilometre Array\footnote{http://www.skatelescope.org/} (SKA). Detection of the redshifted 21-cm signal will not only probe the astrophysics of reionization, but also the matter power spectrum (PS) during the epoch of reionization \citep{McQ+06,B+06}.

In anticipation of forthcoming 21-cm observations, much recent theoretical attention has focused on the prospects of measuring the PS of 21-cm emission \citep[e.g.][]{ZFH04,F+04,morales2005,bowman2006}. In particular, the ionization structure of the intergalactic medium (IGM) owing to UV emission associated with star formation has been studied in detail using analytic~\citep[e.g.][]{F+04,B07} and numerical simulations~\citep[e.g.][]{McQ+06, Il08}. These studies describe a scenario in which very large \H2 regions form around clustered sources within overdense regions of the IGM. The formation of these \H2 regions has a significant effect on the shape of the 21-cm PS, moving power from small to large scales and thus leaving a shoulder-shaped feature on the PS at the characteristic scale of the \H2 regions \citep{F+04}. The detailed morphology of the ionization structure, which is encoded in the PS, will yield information about the ionizing sources, as well as the structure of absorbers in the IGM on small scales~\citep{McQ+06}. More recently, semi-numerical models have also been developed to study the formation of \H2 regions during the epoch of reionization (EoR). These have again focused on the role of UV emission from star-forming regions \citep{MF07} and quasars \citep{GW08}.

In addition to UV photons from the first galaxies and quasars, it has been suggested that a background of X-ray photons at very high redshift may be important in explaining the large Thomson optical depth~\citep{D+08} of the IGM measured by \textit{WMAP} \citep{RO04a,PF07}. Indeed, several authors have interpreted this observation by proposing an initial phase of preheating and partial ionization of the IGM by X-rays, resulting primarily from black hole accretion \citep{SVS85,ME99,HAR00,VGS01,RO04a}. An X-ray background could also have originated from X-ray binaries and supernova remnants, through accretion in the former case and inverse Compton scattering in the latter \citep[see, e.g.][for a review of these processes and the global evolution of the IGM]{FOB06}.

The process by which this early X-ray background ionizes and heats the IGM begins with the photoionization of hydrogen and helium. The resulting high energy free electrons then deposit their energy through collisional ionization and excitation of hydrogen and helium, and electron-electron scattering with cooler free electrons. Thus X-rays ionize hydrogen both directly and through secondary ionizations by photoelectrons from ionized helium, with the latter dominating so that many hydrogen atoms can be ionized by a single X-ray photon. As reionization proceeds, however, the proportion of each X-ray's energy that is deposited into the IGM as heat, increases. In particular, for ionization fractions $\gtrsim 10\%$, ionization by secondary electrons becomes inefficient \citep{RO04a}, as the electron-electron scattering cross-section becomes large. As a result, X-ray ionization is self regulating at the level of $\sim10-20\%$. Further details of the X-ray emission mechanisms are discussed in \cite{O01}. In addition, observational limits on the possible contribution of X-rays to reionization are provided by the unresolved soft X-ray background. \cite{D+04} show that the hard X-ray background produced by a population of black holes at high redshift would be observed as a present-day soft X-ray background. They find that accreting black holes with a hard spectrum could not reionize the Universe to greater than the 50\% level without saturating the unresolved component of the soft X-ray background. 

\cite{PF07} study the effect of the inhomogeneous preheating of the IGM \citep{O01,VGS01,MRVHO04} that results from fluctuations in the X-ray background, and calculate its impact on 21-cm fluctuations at $z\geqslant 10$. The effect of this X-ray heating has also been calculated within numerical simulations~\citep{S+07}. Our work is distinct from these studies, which do not include an X-ray component of reionization. In our modelling we neglect temperature fluctuations, and instead concentrate on the possible effect of X-ray ionization late in the reionization era when \H2 regions dominate the ionization structure, and the IGM is already heated above the cosmic microwave background (CMB) temperature.

The \H2 region dominated ionization structure of the IGM seen in numerical simulations forms as a result of the biased clustering of sources, combined with the short mean free path (MFP) of UV photons. In addition to the relatively large amount of energy deposited into the IGM, X-rays differ from UV photons in their effect on the IGM due to their comparatively long MFP. This long MFP means that X-rays provide a weakly fluctuating radiation background, rather than the highly biased and strongly nonlinear ionizing structure thought to arise from UV photons. The ionization structure of the IGM would therefore be modified if X-rays contributed significantly to reionization, because the long X-ray MFPs imply that their ionizing effect is not confined to the vicinity of large scale overdensities in the matter field \citep{O01}. For this reason, the inclusion of an X-ray component of the emission responsible for the reionization of the IGM could modify fluctuations in the 21-cm PS by reducing the influence of strongly biased ionized regions.

In this paper we extend existing analytic and semi-numerical simulations of reionization to include an X-ray contribution to the ionizing flux. The structure of our paper is as follows: we begin by summarising estimates of the relative X-ray background (\S\,\ref{sec:Xflux}) arising from stellar and accreting stellar mass black hole sources. \S\,\ref{sec:model} describes our analytic model of the 21-cm signal arising from the EoR. In \S\,\ref{sec:model_X} we describe how this model is extended to include a flux contribution from X-rays, and present results for the effect of X-ray ionization on the fluctuation statistics. \S\,\ref{sec:model_num_X} summarizes the semi-numerical scheme used to generate ionization maps of the IGM arising from both UV and X-ray emission. Our semi-numerical results pertaining to the impact of X-rays on the 21-cm PS are discussed in \S\,\ref{sec:results}. In \S\,\ref{noise} we consider the sensitivity of low frequency arrays to X-ray signatures in the 21-cm PS.  We summarize our results in \S\,\ref{sec:summ}.

Throughout this paper we adopt a concordance cosmology for a flat $\Lambda$CDM Universe, $(\Omega_{\rm m},\Omega_{\Lambda},\Omega_{\rm b},h,\sigma_8,n) = (0.27,0.73,0.046,0.7,0.8,1)$, consistent with the \textit{WMAP} constraints from \cite{komatsu2008}. All distances are in comoving units unless stated otherwise.

\section{Estimates of the possible contribution of X-rays to Reionization}
\label{sec:Xflux}

In this section we discuss estimates of the X-ray background at high redshift. We begin by estimating the X-ray photon MFP, and then discuss the X-ray emissivity. High redshift star-formation scenarios might lead to two sources of X-rays, which we consider in turn. In particular, we estimate the X-ray flux from these sources relative to UV flux.  Within the context of this paper, the role of each of these estimates is to motivate different input model parameters for the MFP and fractional X-ray contribution to reionization.

\subsection{The X-ray photon mean free path}
\label{subsec:lambda}

As mentioned in the introduction, the long X-ray MFP implies that they reionize the IGM diffusely via a background, rather than locally as for UV photons. The most critical parameter in studies of the X-ray contribution is, therefore, the MFP. For an X-ray photon of energy $E_{\rm X}$ (measured in eV) the MFP in an IGM of mean neutral fraction $\bar{x}_{\rm H\,{\scriptscriptstyle I}}$ is \citep{MRVHO04,FOB06}
\begin{equation}
\label{eq:lambdaX} 
\lambda_{\rm X}\approx 4.9 \bar{x}_{\rm H\,{\scriptscriptstyle I}}^{-1/3}\left(\frac{1+z}{15}\right)^{-2}\left(\frac{E_{\rm X}}{300\,\rm{eV}}\right)^3\,\rm{Mpc}\,.
\end{equation}
The energy dependence in equation (\ref{eq:lambdaX}) means that in order to estimate the MFP of X-rays, we must specify their spectral energy distribution. To estimate the range of possible X-ray contributions to reionization we assume a spectrum of non-thermal origin with a constant flux per logarithmic interval [$\nu F_{\nu}\sim \rm{constant}\,$], such that $F_\nu \propto 1/E_{\rm X}$, which yields
\begin{eqnarray}
\langle\lambda\rangle &\equiv& \int_{E_{\rm min}}^{E_{\rm max}}\lambda_{\rm X} F_\nu dE_{\rm X} \Big/ \int_{E_{\rm min}}^{E_{\rm max}}F_{\nu}dE\nonumber\\
&=&4.9 \bar{x}_{\rm{H}\,{\scriptscriptstyle I}}^{-1/3}\left(\frac{1+z}{15}\right)^{-2}\,{\rm Mpc}\, \times \nonumber\\
& &\int_{E_{\rm min}}^{E_{\rm max}}dE_{\rm X} \frac{\left({E_{\rm X}}/{300\,\rm{eV}}\right)^3}{E_{\rm X}}\Big/  \int_{E_{\rm min}}^{E_{\rm max}}dE_{\rm X}\frac{1}{E_{\rm X}}\,.
\end{eqnarray}
If we were to assume a steeper energy spectrum \citep[as in][for example]{PF07}, the estimated MFP would be smaller.  Of course the estimate also depends on the range of X-ray energies considered. It is important to note that the MFP is necessarily bounded above by the horizon scale, $c/H$, and all MFPs adopted in this paper adhere to this constraint.  As an example, we can consider an energy range $0.25\,{\rm keV} \leq E_{\rm X} \leq 1\,{\rm keV}$. At $z=(7,9)$, and assuming $\bar{x}_{\rm H\,{\scriptscriptstyle I}}^{-1/3}\sim 1$, this yields $\langle\lambda\rangle \approx (153,98)$\,Mpc.

\subsection{Estimates of the X-ray emissivity during reionization based on the properties of local galaxies}
\label{sec:Xfrac}
We next summarize estimates of the X-ray emissivity arising from a range of star-formation related sources (including supernovae remnants and X-ray binaries) based on observations of local galaxies.  Assuming that the correlation between local star-formation rate (SFR) and X-ray luminosity can be extrapolated to high redshift, the X-ray luminosity during reionization is given by \citep{F06b};
\begin{equation}
\label{eq:mfp}
 L_{\rm X} = 3.4\times 10^{40} f_{\rm X}\left(\frac{\rm SFR}{1\,{\rm M}_\odot\,\rm{yr}^{-1}}\right)\rm{erg}\,\rm{s}^{-1}\,,
\end{equation}
where $f_{\rm X}$ is a normalisation factor that parametrizes ignorance of star formation at high redshift. In this expression the SFR can be approximated by
\begin{equation}
\label{SFReqn}
{\rm SFR} \approx f_\ast \frac{dF_{\rm coll}}{da}\frac{H_0 \Omega^{1/2}_{\rm m}}{a^{1/2}}n_{\rm H}(z)m_{\rm H}\,\rm{kg\,s}^{-1}\,,
\end{equation}
where $a=(1+z)^{-1}$ is the cosmic scale factor, $n_{\rm H}(z)$ is the number density of hydrogen at redshift $z$, $m_{\rm H}$ is the atomic mass of hydrogen, $f_\ast$ is the star-formation efficiency, and we have ignored the effect of the cosmological constant (appropriate at high redshift). Combining equations (\ref{eq:mfp}) and (\ref{SFReqn}), multiplying by the Hubble time and dividing by $n_{\rm H}(z)$, we obtain the luminosity per baryon emitted in X-rays
\begin{equation}
\epsilon_{\rm X} = t_{\rm H}(z)L_{\rm X}/n_{\rm H}\,.
\end{equation}
Finally, the mean number of X-rays per baryon $N_{\rm X}$ can be found by averaging the number of X-rays per baryon over all X-ray energies, which yields
\begin{eqnarray}
\nonumber N_{\rm X} &=& \int d\nu\frac{d\epsilon_{\rm X}}{d\nu}\frac{1}{E_{\rm X}}\nonumber\\
 &=& 	\epsilon_{\rm X}\int d\nu\frac{d\epsilon_{\rm X}}{d\nu}\frac{1}{E_{\rm X}}\big/\int d\nu\frac{d\epsilon_{\rm X}}{d\nu}\nonumber\\
 &=&	5.58\times 10^{-2}\,a f_{\rm X}\frac{f_{\ast}}{0.1}\frac{dF_{\rm{coll}}/da}{0.01}\Big<\frac{1}{E_{\rm X}}\Big>\,.
\end{eqnarray}
The factor describing the mean of the inverse energy of X-rays is evaluated using the flux density $F_\nu$, where again we assume $F_\nu \nu \sim {\rm constant}$,
\begin{eqnarray}
\Big<\frac{1}{E_{\rm X}}\Big> &\equiv& \int_{E_{\rm min}}^{E_{\rm max}}\frac{1}{E_{\rm X}} F_\nu dE_{\rm X} \Big/ \int_{E_{\rm min}}^{E_{\rm max}}F_{\nu}dE_{\rm X}\nonumber\\
&=& \frac{1/E_{\rm min}-1/E_{\rm max}}{\ln\left(E_{\rm max}/E_{\rm min}\right)}\,.
\end{eqnarray}
For an energy range $0.25\,{\rm keV} \leq E_{\rm X} \leq 1\,{\rm keV}$, this expression yields $\langle 1/E_{\rm X} \rangle \sim 1/460\,\rm{eV}$.

It can be shown that for a weakly-ionized IGM only about 1/3 of the X-ray energy contributes to ionization \citep{SVS85,CK04}, but that each X-ray photon produces approximately $12$ ionizations \citep{SVS85,VGS01}, mostly arising from secondary ionizations by photoelectrons from ionized helium. As reionization proceeds, however, the proportion of each X-ray's energy that is deposited into the IGM as heat, increases. In particular, for $\bar{x}_{\rm H\,{\scriptscriptstyle II}} \gtrsim 10\%$, ionization by secondary electrons becomes inefficient \citep{RO04a}, as the electron-electron scattering cross-section becomes large. With these considerations in mind we estimate the number of ionizing X-ray photons per baryon to be
\begin{eqnarray}
\label{eq:NXion}
N_{\rm X,ion} &=& 12 N_{\rm X}\nonumber\\
	&\approx& 0.015\, f_{\rm X}\frac{f_{\ast}}{0.1}\frac{dF_{\rm coll}/d(\ln{a})}{0.05}\Big<\frac{460\,\rm{eV}}{E_{\rm X}}\Big>\,.
\end{eqnarray}
We note that as reionization proceeds, equation (\ref{eq:NXion}) increasingly overestimates the X-ray contribution to reionization.

With regard to the role of X-rays in the reionization of the IGM, it is instructive to compare the number of X-ray to UV ionizations. To achieve this we first evaluate the star-formation rate of UV sources (which is the same as for X-rays here since we are only considering X-rays from stellar emission in this subsection) using equation~(\ref{SFReqn}) and hence the number of ionizing photons per baryon
\begin{equation}
\label{eq:NUVion}
N_{\rm UV,ion} = 0.6 \frac{f_{\rm esc}}{0.03}\frac{f_\ast}{0.1}\frac{dF_{\rm coll}/d(\ln{a})}{0.05}\,,
\end{equation}
where $f_{\rm esc}$ is the fraction of ionizing photons that escape from star-forming galaxies. In evaluating equation~(\ref{eq:NUVion}) we have assumed that 4000 photons are produced per baryon incorporated into stars \citep{BL01}. Using equations (\ref{eq:NXion}) and (\ref{eq:NUVion}) we obtain the relative number of X-ray to UV ionizations, which in turn yields the fractional contribution to reionization by X-rays
\begin{equation}
\label{eq:Nratio}
\frac{N_{\rm X,ion}}{N_{\rm UV,ion}} = 0.025 f_{\rm X}\frac{0.03}{f_{\rm esc}}\Big<\frac{460\,\rm{eV}}{E_{\rm X}}\Big>\,.
\end{equation}
As mentioned above, the normalisation factor $f_{\rm X}$ parametrizes the uncertainty in the relation between the spectral energy distribution produced by low and high redshift star formation. Equation~(\ref{eq:Nratio}) suggests that X-rays would need to comprise a large fraction of the energy budget when compared with low redshift star formation ($f_{X}\gg1$) in order for X-rays following directly from star formation to make a substantial contribution to the reionization of the IGM.

\subsection{X-ray background generated by black hole accretion}
\label{subsec:BH}
An alternative means for estimating the X-ray flux is to consider efficient accretion onto pop-III remnant BHs at high redshift \citep[e.g.][]{RO04a}. In terms of X-ray production, accretion is potentially more efficient than star formation. Indeed, several authors have argued that the number of primary and secondary X-ray ionizations per accreted baryon is $N_{\rm X}\sim 10^6$ \citep{VGS01,RO04a,ROG05,SV08}. Once again we assume that each X-ray photon produces $\sim\,12$ ionizations. As a result, the fraction of baryons accreted, $\omega_{\rm BH}$, need only be very small to achieve of order one X-ray ionization per baryon in the IGM. For the purpose of illustration, we follow previous estimates \citep{RO04a} by assuming $\omega_{\rm BH}\sim 10^{-7}\-- 10^{-6}$, with approximately 1/3 of the total X-ray energy contributing to reionization. The number of X-ray ionizations per baryon in the IGM then becomes \citep{SV08}
\begin{eqnarray}
\label{eq:NX_BH}
N_{\rm X,ion,BH} &\approx& 12\left(\frac{0.1 m_{\rm p}c^2}{2.7\times10^{-9}\,\rm{erg}}\Big<\frac{460\,\rm{eV}}{E_{\rm X}}\Big>\right)\omega_{\rm BH}\nonumber\\ 
&\approx& (0.01\-- 0.1\,{\rm ions/baryon})\Big<\frac{460\,\rm{eV}}{E_{\rm X}}\Big>\,,
\end{eqnarray}
where 0.1 in the numerator is the accretion efficiency and the escape fraction and clumping factor are assumed to be unity. Using Equations (\ref{eq:NUVion}) and (\ref{eq:NX_BH}), and assuming that UV photon production is dominated by stellar sources, we can now estimate the contribution to reionization of X-rays produced by black hole accretion relative to UV photons
\begin{eqnarray}
\label{eq:NX_BH_frac}
\frac{N_{\rm X,ion,BH}}{N_{\rm UV,ion}} &\sim& (0.02\-- 0.2)\Big<\frac{460\,\rm{eV}}{E_{\rm X}}\Big>\nonumber\\
	& &\times\frac{0.03}{f_{\rm esc,UV}}\left(\frac{f_\ast}{0.1}\right)^{-1}\left[\frac{dF_{\rm coll}/d(\ln{a})}{0.05}\right]^{-1}.
\end{eqnarray}

Equation (\ref{eq:NX_BH_frac}) demonstrates that the fractional contribution of X-rays to the ionization of the IGM could be as high as 10s of percent during the early stages of reionization.  This estimate of the number of X-ray ionizations per baryon due to accretion is in excess of that expected directly from star formation, and demonstrates the plausibility of a non-negligible X-ray contribution to reionization. As mentioned earlier, one caveat is that theoretical calculations show that the free electrons produced by reionization limit the fraction of the IGM that can be ionized by X-rays alone to $\lesssim 20\%$ \citep{RO04a}, while observations of the soft X-ray background limit the X-ray contribution to reionization to $<50\%$~\citep{D+04}.  It should be noted that in this paper the X-ray fraction refers to the fraction of IGM ionized by X-rays prior to a particular redshift. In principle, this fraction could therefore be as high as 50\% until midway through reionization. 

Based on the considerations summarized above, we are able to identify a plausible  range for the contribution of X-rays to the reionization of hydrogen. In the modelling presented in the remainder of this paper we therefore analyse scenarios for which there is a fractional contribution to the reionization of hydrogen at redshift $z$ of between 0\% and 50\% due to X-rays with mean energies between of 300eV and 460eV. These energies yield MFPs that straddle the range of values for the characteristic bubble scale during the various stages of the reionization era.

\section{Summary of the Model For Reionization}
\label{sec:model}
We model fluctuations in the 21-cm signal from neutral hydrogen during reionization using a semi-numerical scheme following the work of \cite{GW08}, which is based on the analytic model of \cite{WL07}. The analytic model encodes the preference for galaxy formation in overdense regions \citep{MW96}, in concert with the suppression of low mass galaxies in ionized regions resulting from ionization feedback. The robust prediction of this model is that overdense regions reionize first. Individual \H2 bubbles surrounding clustered sources then percolate into less overdense regions, resulting in the eventual overlap of ionized regions throughout the IGM. This behaviour mimics that seen in sophisticated numerical simulations \citep[e.g.][]{MLZDHZ07}. In this paper we consider a model in which the mean IGM is reionized at $z \sim 6$ \citep{F06,GF06,WBFS03}. The model differentiates between star formation in neutral regions, which proceeds in halos above the hydrogen cooling threshold $T_{\rm min}\sim 10^4\,\rm{K}$, and star formation in ionized regions, which is suppressed below a higher threshold $T_{\rm ion}\sim 10^5\,\rm{K}$ due to radiative feedback.  The model predicts the time dependent ionization fraction by mass $Q_{\delta,R}$ of a particular region on a scale $R$ with overdensity $\delta_R$. On scales much greater than the ionising photon MFP, this ionization fraction evolves according to the following differential equation \citep{WL07}
\begin{eqnarray}
\label{eq:dqdt}
\frac{dQ_{\delta,R}}{dt} &=& \frac{N_{\rm ion}}{0.76} \left[ 
		Q_{\delta,R}\frac{dF_{\rm coll}(\delta_R,R,z,M_{\rm ion})}{dt}\right.\nonumber\\
	& & \left.+ (1-Q_{\delta,R})\frac{dF_{\rm coll}(\delta_R,R,z,M_{\rm min})}{dt}\right]\nonumber\\ 
	& &  -\alpha_{\rm B}Cn_{\rm H}^0 \left[1+\delta_R\frac{D(z)}{D(z_{\rm obs})}\right](1+z)^3 Q_{\delta,R}\,,
\end{eqnarray}
where $N_{\rm ion}$ is the number of ionizing photons per baryon in galaxies, $\alpha_{\rm B}$ is the case-B recombination coefficient, $C$ is the clumping factor (taken to be $2$ in the work presented here) and $D(z_{\rm obs})$ is the growth factor between redshift $z$ and the present. The halo masses $M_{\rm min}$ and $M_{\rm ion}$ correspond to virial masses of temperature $T_{\rm min}$ and $T_{\rm ion}$ respectively. The last term in equation (\ref{eq:dqdt}) describes the recombination rate of ionized gas. Ionizing photons are assumed to be produced at a rate proportional to the collapsed fraction of mass $F_{\rm coll}$ in halos above the minimum thresholds in neutral ($M_{\rm min}$) and ionized ($M_{\rm ion}$) regions. In a region of comoving radius $R$ and mean overdensity $\delta_R(z)$, the relevant collapsed fraction is obtained from the extended Press-Schechter \citep{PS74} formalism \citep{B91}
\begin{equation}
\label{eq:Fcol}
 F_{\rm coll}(\delta_R,R,z) = {\rm erfc}\left[\frac{\delta_{\rm c}-\delta_R(z)}
	{\sqrt{2(\sigma^2_{\rm gal}-\sigma_R^2)}}\right]\,.
\end{equation}
From the ionization fraction $Q_{\delta,R}$ we can calculate the corresponding brightness temperature $T$ using
\begin{equation}
T(\delta_R,R) = 22\,{\rm mK}\left(\frac{1+z}{7.5}\right)^{1/2}(1-Q_{\delta,R})(1+\delta_R)\,.
\end{equation}

\section{Extension of analytic model to include X-rays}
\label{sec:model_X}
We extend the model described in \S\,\ref{sec:model} to account for the contribution of X-rays to the reionization of the IGM.  The X-ray contribution is modelled as six coupled differential equations describing the contributions of X-rays and UV photons on scales $R$ ($Q_{X,\delta, \rm R}$ and $Q_{UV,\delta, \rm R}$ respectively) and $\langle\lambda\rangle$ ($Q_{X,\delta, \lambda}$ and $Q_{UV,\delta, \lambda}$ respectively) to the evolution of the ionized fraction  on a scale $R$ ($Q_{\delta, \rm R}$).  Separate rates must be computed because at any point in the IGM i) X-rays and UV photons cause different levels of radiative feedback, ii) they are generated from regions of IGM with different scales and overdensities, and iii) the relative contribution to reionization from UV and X-ray photons is sensitive to radiative feedback on different scales.  The equations are:
\begin{equation}
\label{eq:dqdt_X}
\frac{dQ_{\delta,R}}{dt} = \frac{dQ_{UV,\delta,R}}{dt}+\frac{dQ_{X,\delta,\lambda}}{dt}\,,
\end{equation}
\begin{equation}
\frac{dQ_{\delta,\lambda}}{dt} = \frac{dQ_{UV,\delta,\lambda}}{dt}+\frac{dQ_{X,\delta,\lambda}}{dt}\,,
\end{equation}

\begin{eqnarray}
\frac{dQ_{UV,\delta,R}}{dt} &=& (1-X_{\rm frac})\frac{N_{\rm ion}}{0.76} \times \nonumber\\
	& &\hspace{-18mm}\left\lbrace\mathcal{H}(Q_{\rm{crit}}-Q_{X,\delta,R})\left[Q_{\delta,R}\frac{dF_{\rm coll}(\delta_R,R,z,M_{\rm ion})}{dt} \right.\right.\nonumber\\ 
	& &\hspace{-18mm} \left. + (1-Q_{\delta,R})\frac{dF_{\rm{coll}}(\delta_R,R,z,M_{\rm min}) } {dt}\right]\nonumber\\ 
	& & \hspace{-18mm}\left. + \mathcal{H}(Q_{X,\delta,R}-Q_{\rm{crit}}) \left[\frac{dF_{\rm coll}(\delta_R,R,z,M_{\rm ion})}{dt}\right]\right\rbrace\nonumber\\ 
	& & \hspace{-18mm}+B_{\delta,R}\frac{Q_{\rm{UV},\delta,R}}{Q_{\delta,R}}\nonumber\,,\\ 
\label{eq:QUVR}
\end{eqnarray}
\begin{eqnarray}
\frac{dQ_{UV,\delta,\lambda}}{dt} &=& (1-X_{\rm frac})\frac{N_{\rm ion}}{0.76} \times \nonumber\\ 	
	& &\hspace{-18mm}\left\lbrace\mathcal{H}(Q_{\rm{crit}}-Q_{X,\delta,\lambda})\left[Q_{\delta,\lambda}\frac{dF_{\rm{coll}}(\delta_{\langle\lambda\rangle},\langle\lambda\rangle,z,M_{\rm ion})}{dt}\right.\right.\nonumber\\
	& &\hspace{-18mm} \left. + (1-Q_{\delta,\lambda})\frac{dF_{\rm{coll}}(\delta_{\langle\lambda\rangle},\langle\lambda\rangle,z,M_{\rm min})}{dt} \right]\nonumber\\
	& &\hspace{-18mm} \left.-\mathcal{H}(Q_{X,\delta,\lambda}-Q_{\rm{crit}})\left[\frac{dF_{\rm{coll}}(\delta_{\langle\lambda\rangle},\langle\lambda\rangle,z,M_{\rm ion})}{dt}\nonumber\right]\right\rbrace\nonumber\\
	& &\hspace{-18mm}+B_{\delta,\lambda}\frac{Q_{\rm{UV},\delta,\lambda}}{Q_{\delta,\lambda}}\nonumber\,\\
\label{eq:QUVlam}
\end{eqnarray}
\begin{eqnarray}
\frac{dQ_{X,\delta,R}}{dt} &=& X_{\rm frac}\frac{N_{\rm ion}}{0.76} \left(\frac{1+\delta_{\lambda}}{1+\delta_{\rm R}}\right)\times \nonumber\\ 	
	& &\hspace{-18mm}\left\lbrace\mathcal{H}(Q_{\rm{crit}}-Q_{X,\delta,R})\left[Q_{\delta,R}\frac{dF_{\rm{coll}}(\delta_{\langle\lambda\rangle},\langle\lambda\rangle,z,M_{\rm ion})}{dt}\right.\right.\nonumber\\
	& & \hspace{-18mm}\left. + (1-Q_{\delta,R})\frac{dF_{\rm{coll}}(\delta_{\langle\lambda\rangle},\langle\lambda\rangle,z,M_{\rm min})}{dt} \right]\nonumber\\
	& &\hspace{-18mm} \left.-\mathcal{H}(Q_{X,\delta,R}-Q_{\rm{crit}})\left[\frac{dF_{\rm{coll}}(\delta_{\langle\lambda\rangle},\langle\lambda\rangle,z,M_{\rm ion})}{dt}\nonumber\right]\right\rbrace\nonumber\\
	& &\hspace{-18mm}+B_{\delta,R}\frac{Q_{\rm{X},\delta,R}}{Q_{\delta,R}}\nonumber\,,\\
\label{eq:QXR}
\end{eqnarray}
\begin{eqnarray}
\frac{dQ_{X,\delta,\lambda}}{dt} &=& X_{\rm frac}\frac{N_{\rm ion}}{0.76} \times \nonumber\\ 	
	& &\hspace{-18mm}\left\lbrace\mathcal{H}(Q_{\rm{crit}}-Q_{X,\delta,\lambda})\left[Q_{\delta,\lambda}\frac{dF_{\rm{coll}}(\delta_{\langle\lambda\rangle},\langle\lambda\rangle,z,M_{\rm ion})}{dt}\right.\right.\nonumber\\
	& &\hspace{-18mm} \left. + (1-Q_{\delta,\lambda})\frac{dF_{\rm{coll,X}}(\delta_{\langle\lambda\rangle},\langle\lambda\rangle,z,M_{\rm min})}{dt} \right]\nonumber\\
	& &\hspace{-18mm} \left.-\mathcal{H}(Q_{X,\delta,\lambda}-Q_{\rm{crit}})\left[\frac{dF_{\rm{coll}}(\delta_{\langle\lambda\rangle},\langle\lambda\rangle,z,M_{\rm ion})}{dt}\nonumber\right]\right\rbrace\nonumber\\
	& &\hspace{-18mm}+B_{\delta,\lambda}\frac{Q_{\rm{X},\delta,\lambda}}{Q_{\delta,\lambda}}\nonumber\,,\\
\label{eq:QXlam}
\end{eqnarray}
where the recombination rates on scales $R$ and $\lambda$, $B_{\delta,R}$ and $ B_{\delta,\lambda}$ are given by
\begin{equation}
 B_{\delta,R} =  -\alpha_{\rm B}Cn_{\rm H}^0\left[1+\delta_R\frac{D(z)}{D(z_{\rm obs})}\right](1+z)^3 Q_{\delta,R}\,,
\label{eq:BR}
\end{equation}
and
\begin{equation}
 B_{\delta,\lambda} =  -\alpha_{\rm B}Cn_{\rm H}^0\left[1+\delta_\lambda\frac{D(z)}{D(z_{\rm obs})}\right](1+z)^3 Q_{\delta,\lambda}\,.
\label{eq:Blam}
\end{equation}
where $X_{\rm frac}$ is the fraction of the total number of ionizations that are due to X-rays as estimated in \S\,\ref{sec:Xfrac}.  Throughout this paper $N_{\rm{ion}}$ is chosen such that the mean ionization fraction at $z=7$ is 0.7 and at $z=9$ is 0.3, which, in the no-X-ray case, corresponds to complete reionization at $z=6$.  $\mathcal{H}$ is the Heaviside step function, which is equal to one when the argument is positive definite, and zero otherwise.  $F_{\rm{coll}}$ is given by equation (\ref{eq:Fcol}).  The term in equation (\ref{eq:QXR}) that is proportional to the ratio of densities on scales $\lambda$ and $R$ describes the ionization of a region of overdensity $\delta_{\rm R}$ due to X-ray photons generated in a region of overdensity $\delta_{\lambda}$.

In this paper we consider, for simplicity, only the mean scale $\langle\lambda\rangle$, rather than the full distribution of MFPs for X-rays of different energies.  Evaluation of $F_{\rm{coll}}(\delta_{\langle\lambda\rangle},\langle\lambda\rangle,z,M_{\rm ion})$ requires calculation of the correlation between $\delta_{R}$ and $\delta_{\langle \lambda\rangle}$, which are concentric. This correlation is implicit in the application of equation~(\ref{eq:Fcol}) to the semi-numerical model in \S~\ref{sec:model_num_X}. However, in our analytic calculations we make the further simplifying assumption that the fraction of matter that has collapsed into halos and that contributes to X-ray ionization in a region of radius $R$ depends on the overdensity as 
\begin{equation}
F_{\rm{coll,\rm R}} = \left\{\begin{array}{ll} 
F_{\rm{coll}} (\delta_R,R,z) & R>\langle\lambda\rangle \vspace{1mm} \\
(R^3/\langle\lambda\rangle^3)F_{\rm coll}(\delta_R,R,z)\ + & \\
(1-R^3/\langle\lambda\rangle^3)F_{\rm coll} (\delta_0,\infty ,z) & R< \langle\lambda\rangle
\end{array}
\right.\,,
\label{eq:FcollX}
\end{equation}
where $\delta_0\equiv 0$ is the mean overdensity of the Universe.  Equation~(\ref{eq:FcollX}) describes the difference imprinted on large versus small scale fluctuations by the addition of X-rays. Firstly, if $\langle\lambda\rangle < R$ then all photons, X-ray and UV, are generated inside the region where the fluctuation is measured (i.e. on a scale $R$).  When $\langle\lambda\rangle > R$, UV photons are generated inside the region $R$, but X-rays come from both inside $R$ and from the region bounded by $R$ and $\langle\lambda\rangle$.  In the latter case, X-rays produced inside $R$ are subject to the same bias as are UV photons; however X-rays from beyond $R$ are produced in a region of different overdensity.  In our analytic calculations we assume for simplicity that X-rays from beyond $R$ are generated at mean density (as already mentioned, this assumption is relaxed in \S~\ref{sec:model_num_X}).

As discussed in \S\,\ref{sec:Xfrac}, approximately 2/3 of the total energy emitted in X-rays contributes to heating the IGM, and the remaining 1/3 contributes to ionization.    As the IGM temperature increases, the formation of smaller mass galaxies is suppressed, which should be reflected in the minimum scale on which the density field is filtered.  We account for the effects of X-ray heating by monitoring the temperature change of the IGM that is due to X-rays, which we assume to be proportional to the ionization fraction due to X-rays.  When the X-ray ionization fraction reaches the heating threshold, $Q_{\rm{crit}}$, defined by
\begin{equation}
Q_{\rm{crit}} = 0.03\frac{T_{\rm{IGM}}}{10^4}\frac{\eta}{2}\,\rm{,}
\end{equation} 
we increase the minimum scale at which the density is filtered from $T_{\rm{min}}$ to $T_{\rm{ion}}$.  $T_{\rm{IGM}}\sim 10^4\,\rm{K}$ is taken as the characteristic IGM temperature above which galaxy formation is suppressed.  We take $Q_{\rm{crit}} = 0.03$ to be the fiducial value and use $\eta$ to parametrize our ignorance of the X-ray spectrum, as well as the emissivity history.  We take $T_{\rm ion}=10^5\,\rm{K}$ to be the fiducial value but allow it to vary given the uncertainty in $T_{\rm ion}$ when $T_{\rm IGM}=10^5\,\rm {K}$.  This physics is included in our model via the Heaviside step functions in equations \ref{eq:QUVR}\,--\,\ref{eq:QXlam}.

The simple analytic model described above can be used to qualitatively demonstrate the impact of X-rays on the statistics of brightness fluctuations during reionization. We describe these results in the following subsection before moving on to more realistic calculations based on our semi-numerical model. We emphasise that throughout this paper we assume that the IGM is considerably hotter than the CMB, and hence do not consider temperature fluctuations as additional sources of 21-cm fluctuations.

\subsection{Results - analytic}
\label{sec:anal_res}

\begin{figure*}
\vspace{1cm}
\includegraphics[width = 6.2cm,angle=90]{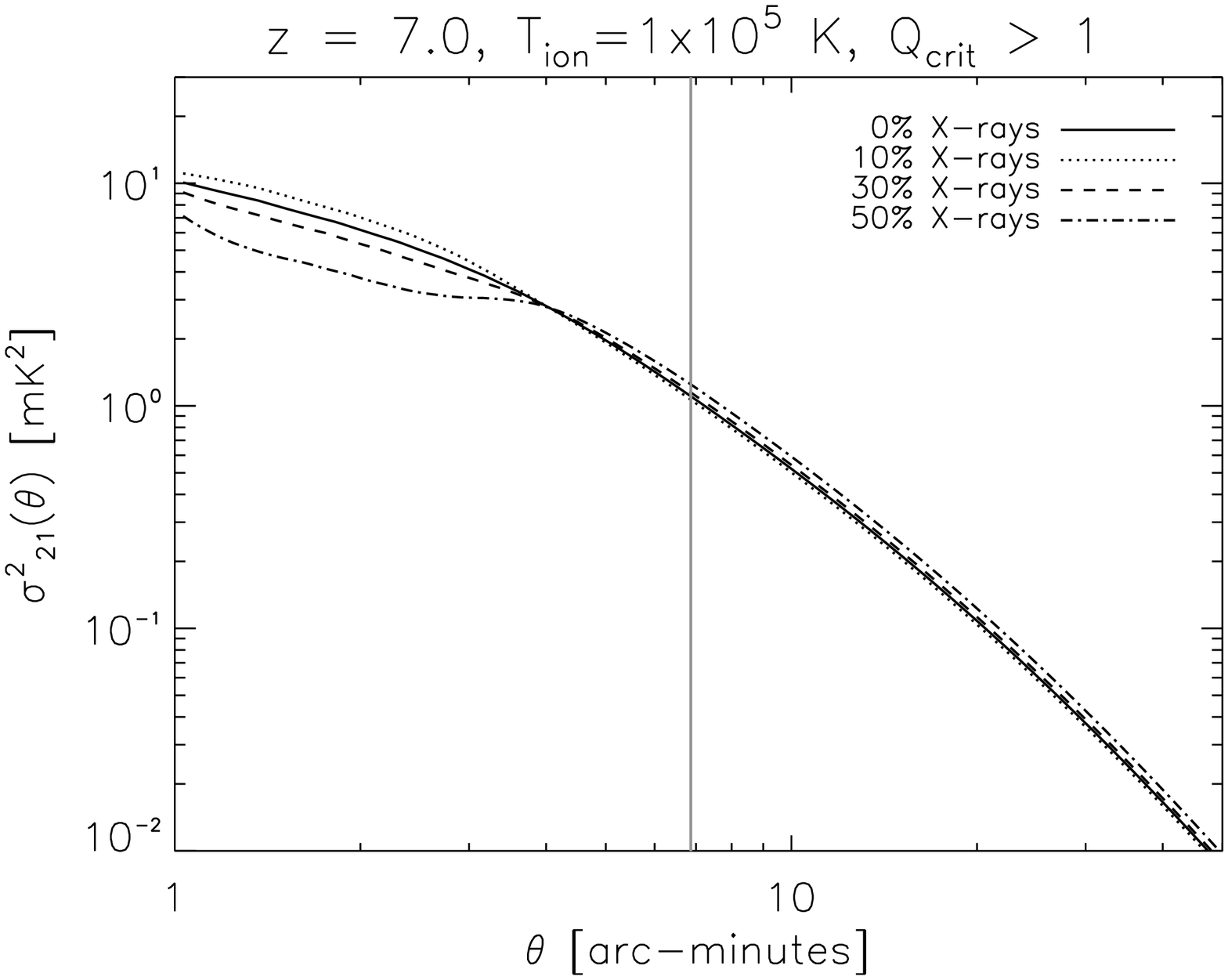}
\includegraphics[width = 6.2cm,angle=90]{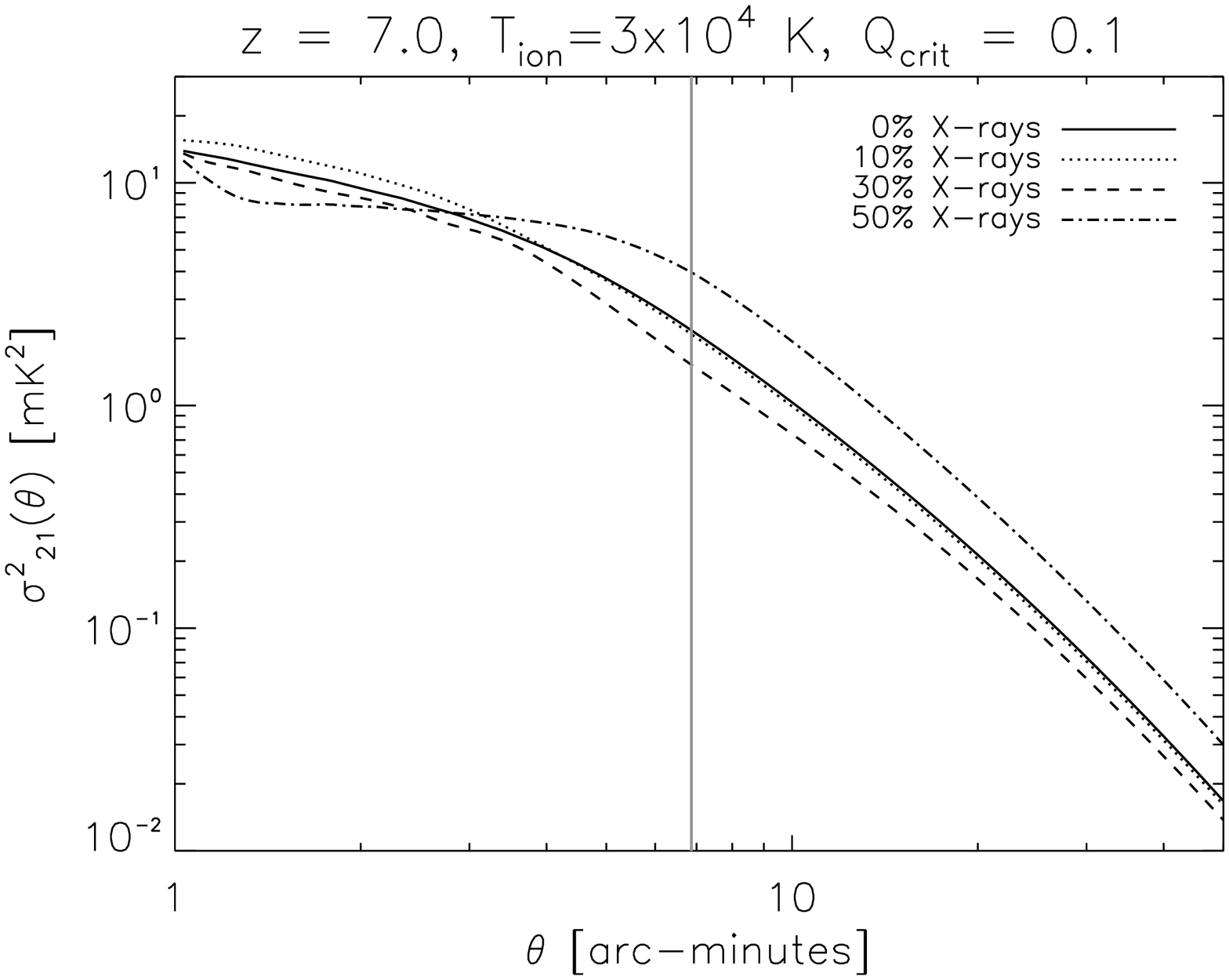}
\includegraphics[width = 6.2cm,angle=90]{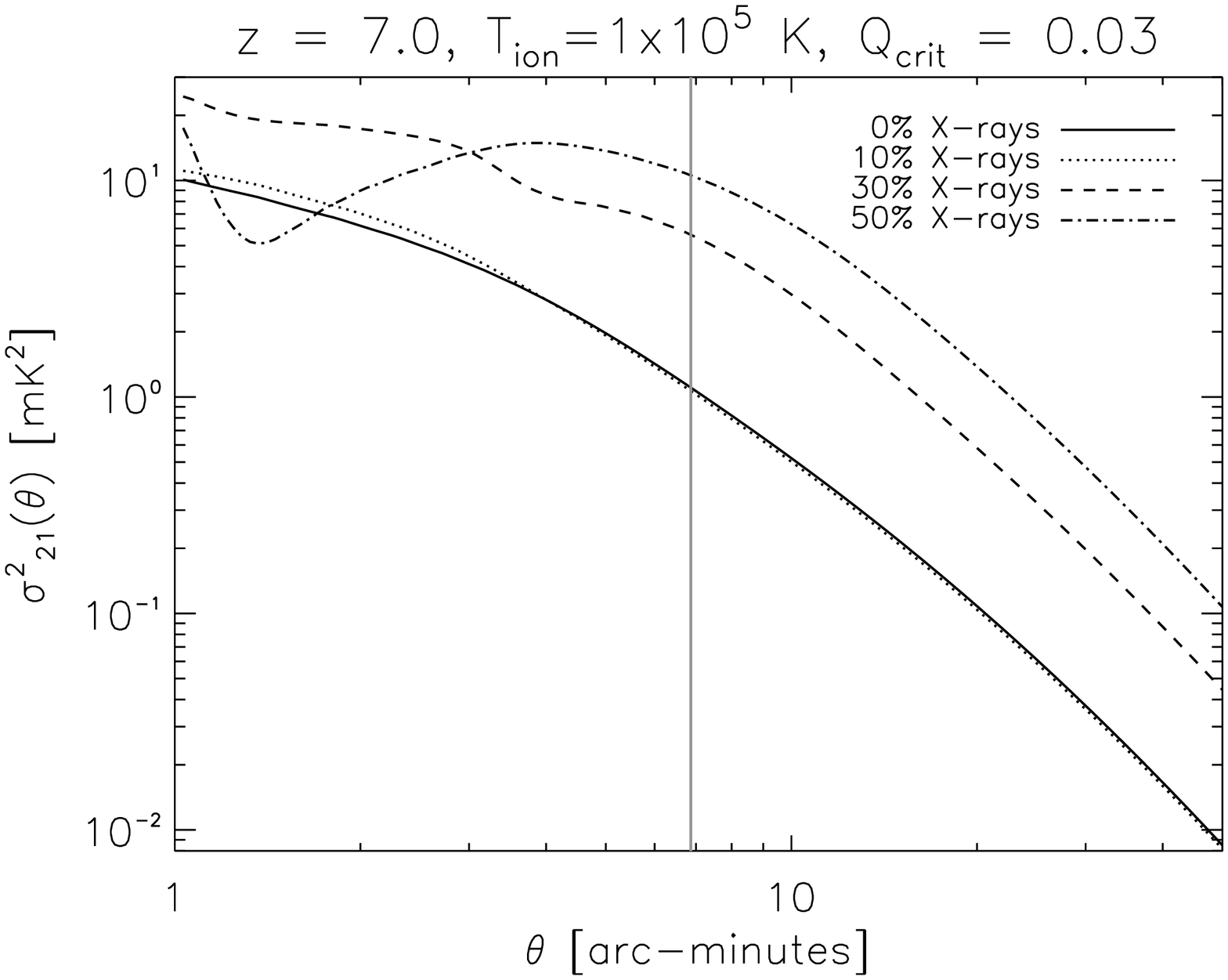}
\includegraphics[width = 6.2cm,angle=90]{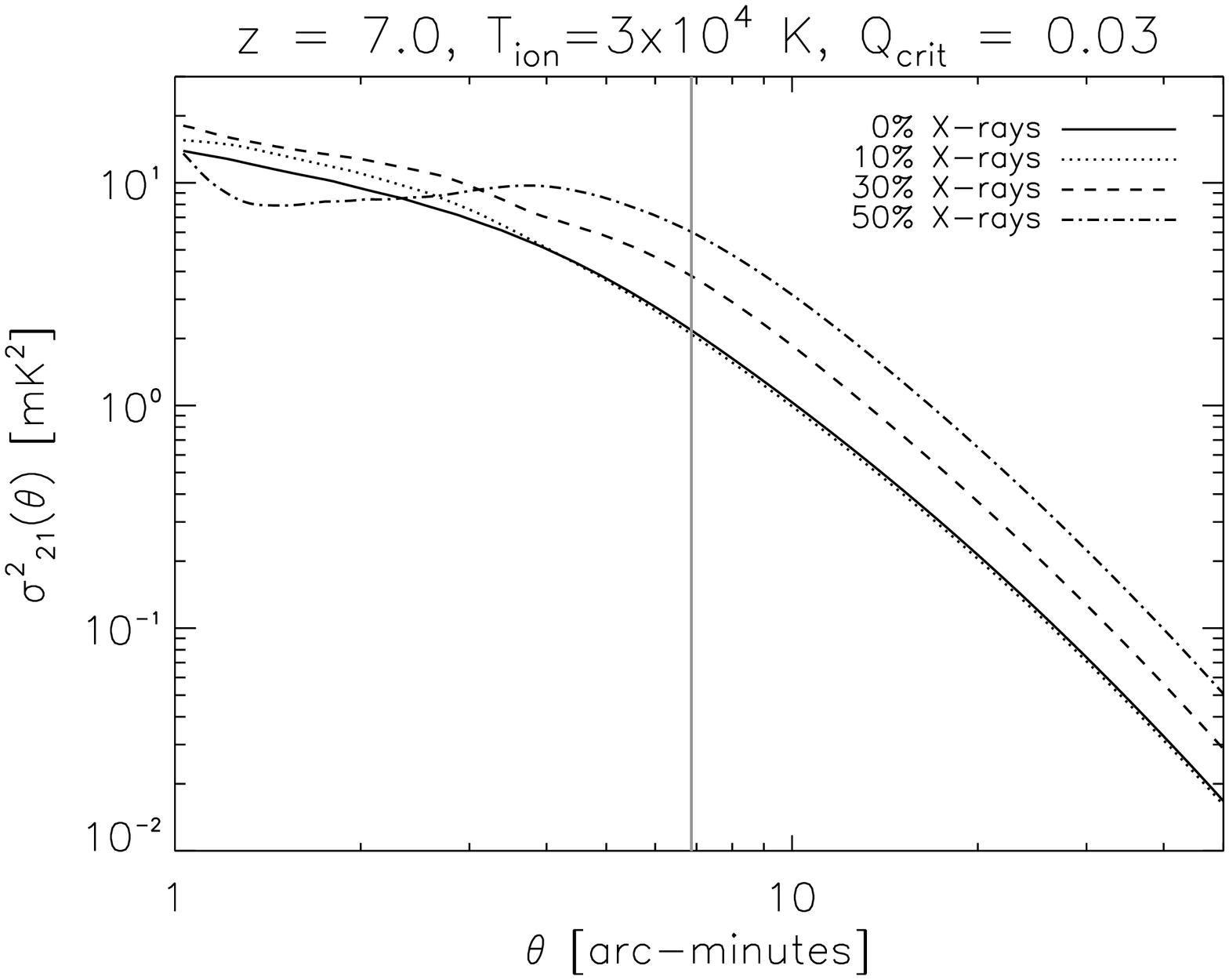}
\caption{The variance of the 21-cm temperature brightness as a function of angular scale at $z=7$.  The X-ray energy is taken to be 300\,eV and X-ray contributions are considered at the 0\% (\emph{solid}), 10\% (\emph{dotted}), 30\% (\emph{dash}) and 50\% (\emph{dash dot}) level.  The vertical lines (\emph{grey}) mark the angular scale of the MFP of 300\,eV X-rays, which is 7\,arc-minutes.  }
\label{fig:analytic_7}
\end{figure*}

\begin{figure*}
\vspace{1cm}
\includegraphics[width = 6.2cm,angle=90]{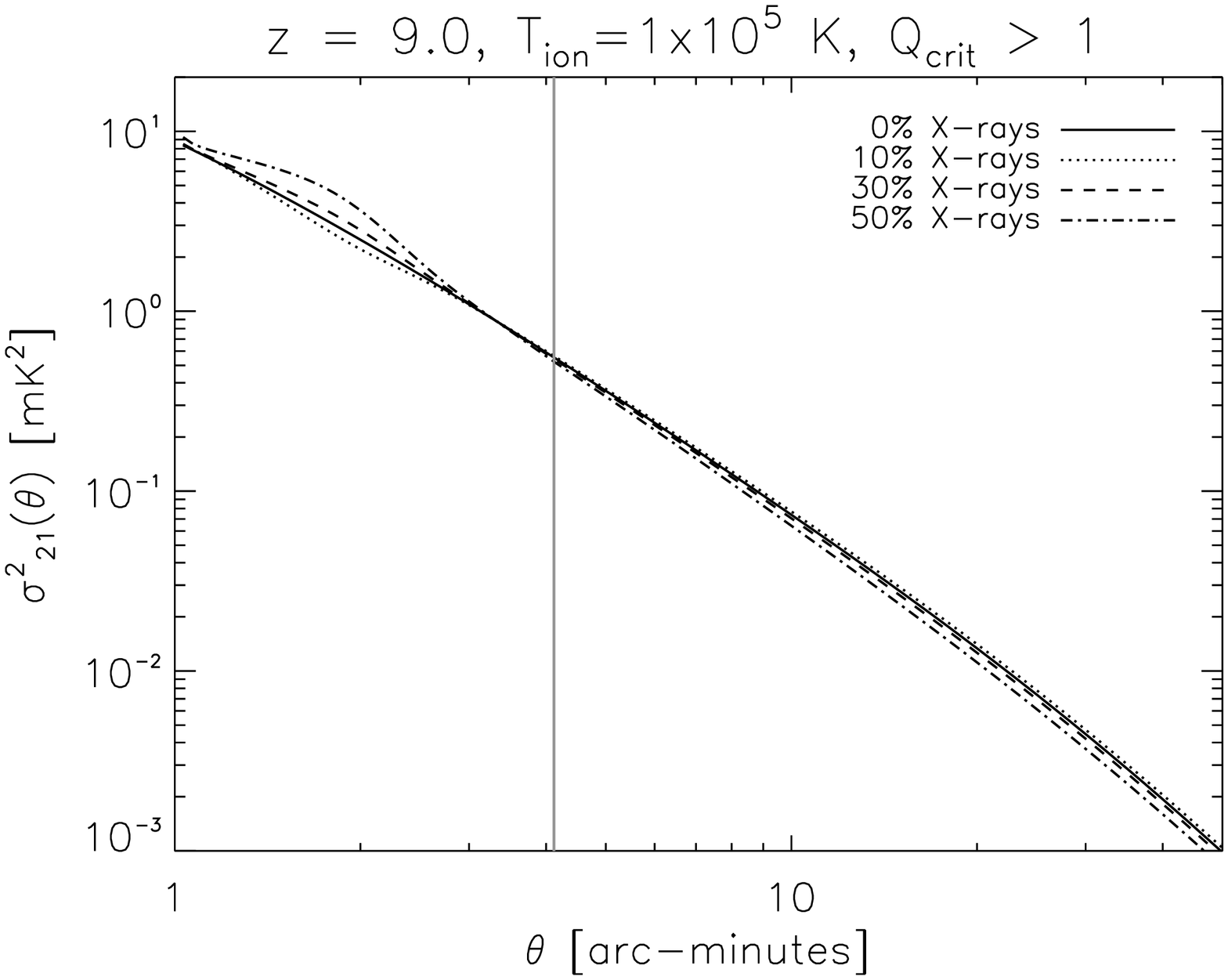}
\includegraphics[width = 6.2cm,angle=90]{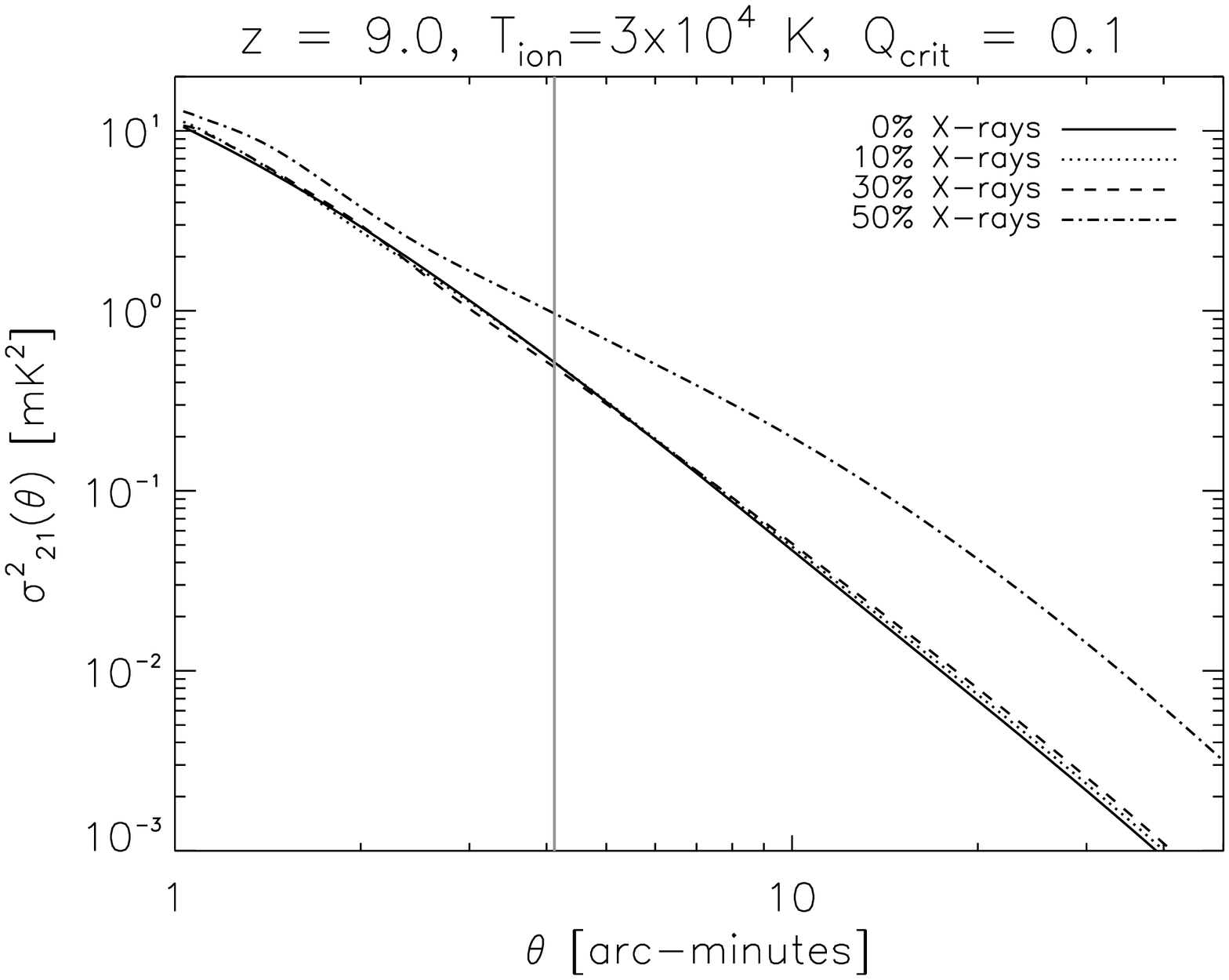}
\includegraphics[width = 6.2cm,angle=90]{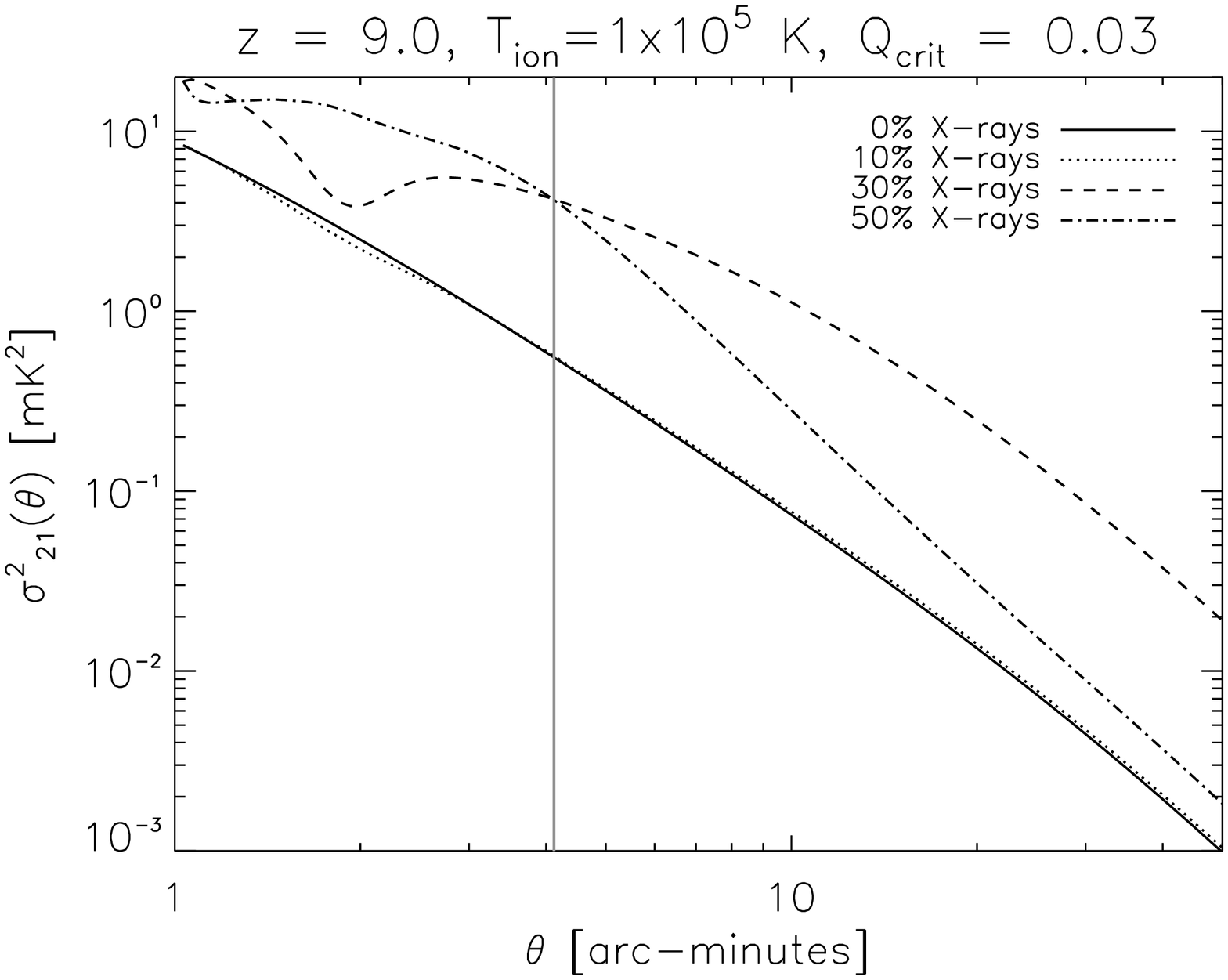}
\includegraphics[width = 6.2cm,angle=90]{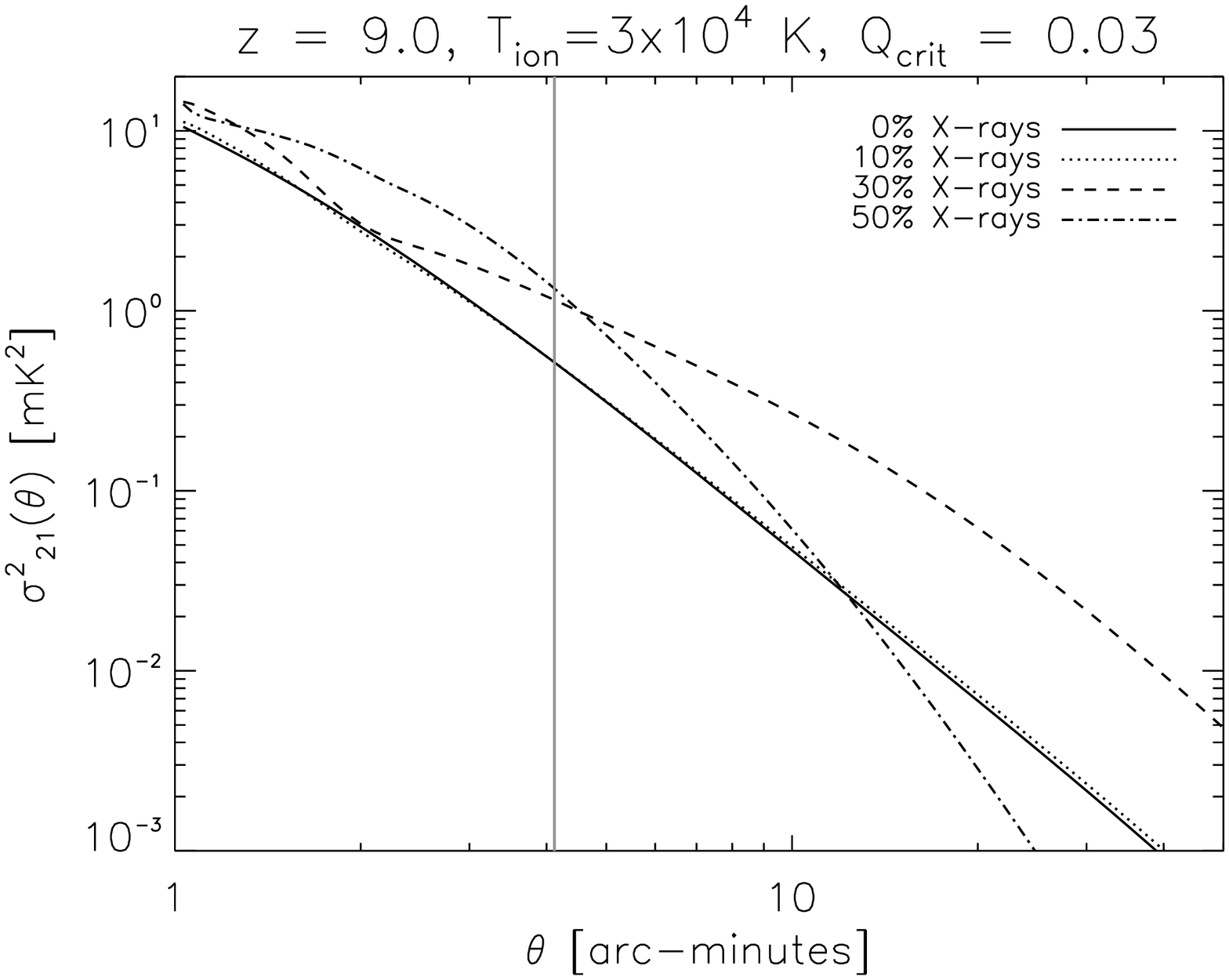}
\caption{The variance of the 21-cm temperature brightness as a function of angular scale at $z=9$ (\emph{right}).  The X-ray energy is taken to be 300\,eV and X-ray contributions are considered at the 0\% (\emph{solid}), 10\% (\emph{dotted}), 30\% (\emph{dash}), and 50\% (\emph{dash dot}) level.  The vertical lines (\emph{grey}) mark the angular scale of the MFP of 300\,eV X-rays, which is 4\,arc-minutes at $z=9$.  }
\label{fig:analytic_9}
\end{figure*}

We express results from our analytic calculation via the variance of 21-cm brightness temperature within regions of physical size corresponding to an observed angle $\theta$. This is given by 
\begin{eqnarray}
\nonumber
\sigma^2_{21}&\equiv&\langle (T-\langle T\rangle)^2\rangle\\
 &=& \frac{1}{\sqrt{2\pi\sigma_R^2}} \int d\delta\,(T(\delta,R)-\langle T\rangle)^2 e^{-\delta^2/2\sigma_R^2}\,,
\end{eqnarray}
where
\begin{equation}
\langle T\rangle = \frac{1}{\sqrt{2\pi\sigma_R^2}}\int d\delta\,T(\delta,R)e^{-\delta^2/2\sigma_R^2}\,.
\end{equation}
Figures \ref{fig:analytic_7} and \ref{fig:analytic_9} show $\sigma_{21}^{2}(\theta)$ as a function of angular scale at $z=7$ and $z=9$ respectively. In this example, the mean X-ray energy is taken to be 300\,eV and fractional X-ray contributions are shown at the 0\%, 10\%, 30\% and 50\% level.  We reiterate that these contributions reflect the fraction of the aggregate ionizing radiation (UV and X-ray photons) that are X-rays. 

We first consider the case in which X-ray heating has a  negligible effect on galaxy formation ($Q_{\rm{crit}}>1$).  The results for this fiducial case are shown in the upper left panels of Figures \ref{fig:analytic_7} and \ref{fig:analytic_9}, where it can be seen that the variance of the 21-cm signal is insensitive to the X-ray contribution for all angular scales $\theta >\theta_{\langle\lambda\rangle}$ (where $\theta_{\langle\lambda\rangle}$ is the angular scale corresponding to the MFP).  On scales $\theta <\theta_{\langle\lambda\rangle}$ the X-ray contribution alters the variance, with the sign of the effect depending on the epoch.  At $z=7$, (for the chosen model) the variance on scales less than $\theta_{\langle\lambda\rangle}$ is reduced relative to the no-X-ray scenario.  This is because the presence of X-rays lessens the bias of 21-cm emission toward underdense regions (which results from reionization by UV photons that are restricted by galaxy bias and short MFP to overdense regions of the IGM).  It should be noted that the analytic calculations cannot describe the effect of ionized bubbles on $\sigma_{21}^{2}(\theta)$, which imprint a characteristic, redshift-dependent scale on the PS.  At $z=9$ the effect of X-rays on $\sigma_{21}^{2}(\theta)$ is reversed:  fluctuations on scales below $\theta_{\langle\lambda\rangle}$ are enhanced by the presence of X-rays.  

\begin{figure*}
\includegraphics[width = 13cm]{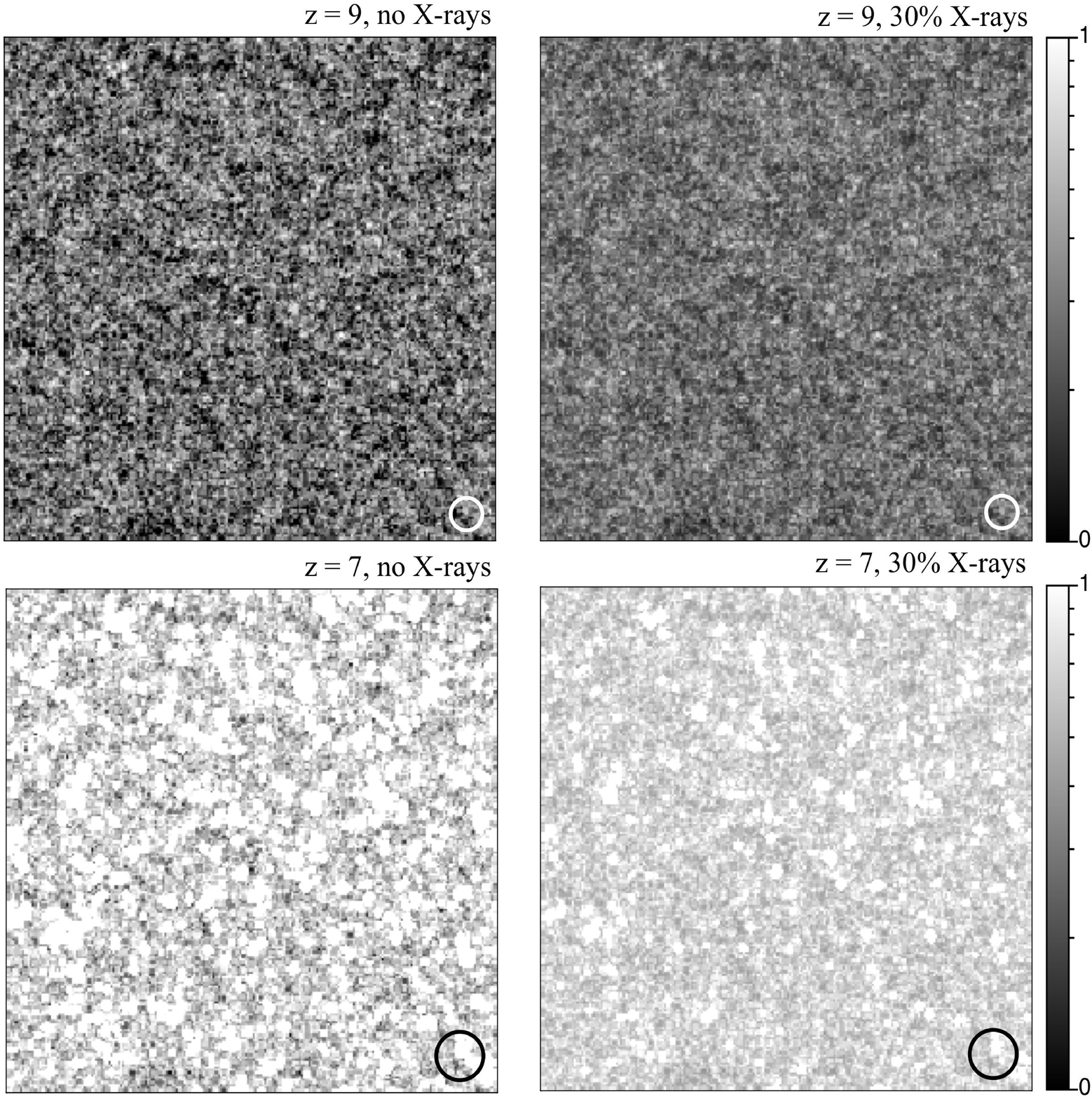}
\caption{Slices from 3D ionization maps at redshifts $9$ (\emph{top}) and $7$ (\emph{bottom}) for the fiducial case with no X-ray contribution (\emph{left}) and 30\% X-ray contribution (\emph{right}).  The colour scale ranges from neutral (\emph{black}) to ionized (\emph{white}).  The circles in the bottom right corner represent the scale of the assumed MFP for 460\,eV X-rays. The simulation slices are 1800\,cMpc on a side.}
\label{fig:Q_maps}
\end{figure*}

These trends can be explained by noting that when the average ionization fraction in the IGM is $\langle Q\rangle \lesssim 0.5$, overdense regions have a higher brightness temperature than underdense regions \citep{WM07}. For our simulations, $Q$ in an overdense region is smaller when X-rays are present because the X-ray contribution is proportional to the mean overdensity, and hence is reduced relative to the UV contribution in an overdense region. Since brightness temperature increases as $Q$ decreases, we expect to see a relative enhancement in brightness temperature in overdense regions on scales where the X-ray contribution to reionization differs from the UV contribution (i.e. $\theta < \theta_{\langle\lambda\rangle}$).  Similarly, in underdense regions $Q$ is greater when X-rays are present and hence the brightness temperature is smaller. The combined effect of these two factors is to enhance fluctuations in the temperature brightness if $\langle Q\rangle\lesssim 0.5$. When $\langle Q\rangle\gtrsim 0.5$ the converse holds: on small scales, $\theta <\theta_{\langle \lambda\rangle}$, and relative to the mean temperature brightness of the IGM, X-rays suppress fluctuations by boosting the ionization fraction in underdense regions and reducing it in overdense regions. The properties of the analytic model when X-ray heating is ignored do not vary qualitatively with X-ray energy or fractional X-ray contribution.

The inclusion of X-ray heating of the IGM complicates the interpretation of the resulting 21-cm variance.  Most notably, there is no longer quantitative agreement between the variance arising from different X-ray fractions on scales greater than $\theta_{\langle\lambda\rangle}$.  X-ray heating suppresses the formation of small mass galaxies, which increases the clustering bias of emission sources.  The greater the bias of sources, the earlier one expects overdense regions to be ionized relative to underdense regions.  There are, therefore, two competing effects due to X-rays:  the tendency for X-rays to shift ionizations from overdense to underdense regions due to their long MFP, and the increase in source bias due to the suppression of low mass galaxies.  The upper right and lower panels of figures \ref{fig:analytic_7} and \ref{fig:analytic_9} explore the interplay of these two effects by varying $Q_{\rm{crit}}$ and $T_{\rm{ion}}$.  In the lower left panels, $Q_{\rm{crit}}$ is decreased to $0.03$, such that regions of the IGM that have been heated by X-rays to at least $10^4\,\rm{K}$ are filtered on the same mass scale as fully ionized regions.  The most notable result of reducing $Q_{\rm{crit}}$ at $z=7$ is that the suppression of fluctuations on scales smaller than $\theta_{\langle \lambda\rangle}$ is not as prominent as when X-ray heating is ignored.  At $z=9$ this effect is more dramatic, as X-ray heating acts to suppress, rather than enhance 21-cm fluctuations.  In addition, the results in the panels of figures \ref{fig:analytic_7} and \ref{fig:analytic_9} in which heating has been considered expose a trend for fluctuations to increase overall as the X-ray fraction is increased.  We owe this effect to the precocious ionization of overdense regions, and hence of a greater fraction of the total hydrogen content of the IGM, due to an increase in the bias of emission sources.  As expected, the effect is more pronounced at high redshift compared to epochs closer to overlap.

If we decrease $T_{\rm{ion}}$ to $3\times 10^4\,\rm{K}$ (lower right panel of figures \ref{fig:analytic_7} and \ref{fig:analytic_9}), with $Q_{\rm{crit}}=0.03$, the impact of including X-ray heating is reduced overall, as the filtering of ionized regions is less distinct from that of neutral regions.  Finally, if we increase $Q_{\rm{crit}}$ to 0.1 (upper right panels of figures \ref{fig:analytic_7} and \ref{fig:analytic_9}), the impact of X-ray heating varies depending on the fractional X-ray contribution to reionization.  For small X-ray contributions ($\sim 10\%$), the shift in bias to overdense regions is the dominant effect of X-ray heating.  As the X-ray contribution increases, however, the suppression of fluctuations on scales smaller than $\theta_{\langle\lambda\rangle}$ becomes more significant.

\section{Inclusion of X-rays in a Semi-numerical scheme}
\label{sec:model_num_X}

Several recent studies of the 21-cm signal during the EoR have employed semi-numerical techniques to reduce the computational cost of large-scale numerical simulations \citep{MF07,ZLMDHZF07,GW08,CHR08}. The advantage of semi-numerical simulations, as compared to analytic models, is the ability to trace the formation and overlap of ionized bubbles, which feature prominently in the PS.  On the other hand, semi-numerical techniques do not capture variations in the actual halo counts from linear theory. The modelling of \cite{MLZDHZ07} and \cite{Z+07} shows this to be a significant effect.

The semi-numerical model in this paper is based on the method implemented by \cite{GW08}. It employs a three-dimensional realisation of a Gaussian random field, representing fluctuations in the matter overdensity.  The simulation is performed in a $200^3$ box, with side length equal to 400 and 1800\,cMpc for X-rays with mean energies of $300$ and $460$ eV (for a flat X-ray spectrum) respectively, with an underlying $\Lambda$CDM PS \citep{EH98} evolved linearly to the specified redshift $z$. The density field is then smoothed repeatedly on scales ranging from the pixel size to the size of the simulation box. At each scale we calculate the ionized hydrogen fraction $Q$ according to equation (\ref{eq:dqdt_X}). The calculation of $Q$ differs from the analytic case in that the overdensity on the scale of $\langle\lambda\rangle$ is calculated explicitly, rather than assuming it to be equal to the mean overdensity. For this reason, there is no longer a need to differentiate between smoothing scales greater and less than $\langle\lambda\rangle$. An ionized spherical bubble of radius $R$ forms around voxels in which $Q_{\delta,R}\geq 1$, producing a two-phase ($Q=0$ or $1$) ionization map of the IGM. Voxels that do not form part of an ionized bubble after the field has been smoothed on all scales, are then assigned a partial ionization fraction corresponding to the value of $Q$ obtained by smoothing the density field on the minimum (pixel) scale. Of importance to this study, which requires very large box sizes, is that the use of equations~(\ref{eq:dqdt_X})--(\ref{eq:Blam}) rather than a barrier formalism \citep[e.g.][]{MF07} means that partially ionized pixels are treated correctly with respect to the ionization contribution of unresolved \H2 regions.  

It should be noted that the ionization field is simulated at a fixed proper time, and as a result the corresponding 21-cm PS is computed assuming a non-evolving IGM. However, in practice, a single observation measures the redshifted 21-cm signal over a range of cosmic time corresponding to the particular frequency band of the observation. The consequence of this is that the observed spherically averaged 21-cm PS is modified on the largest scales, because the ionization state of the IGM evolves significantly during the light travel time corresponding to the frequency band.  We do not address this evolution in the present study. 

Physically, two principle features are captured in the semi-numerical scheme that are not present in our semi-analytic formulation. Firstly, the presence of ionized regions, which imprint a scale dependent feature on the PS. And secondly, the density dependent calculation of X-ray contribution.

\section{Results - semi-numerical}
\label{sec:results}

In order to examine the combined impact of \H2 regions and X-ray ionization on the statistics of 21-cm fluctuations, we now consider the results of the semi-numerical realisation of the model outlined in \S\,\ref{sec:model}. We explore the parameter space of average X-ray MFP $\langle \lambda\rangle$ and the fraction of ionizations due to X-rays (reflecting the uncertainty in $f_{\rm X}$, $\langle \lambda\rangle$ and $f_{\rm esc}$) at $z=7$ and $9$.  The chosen redshifts straddle the epoch when the mass-weighted, average ionization fraction $\langle Q\rangle_{\rm mass} \sim 0.5$ for this model.  We consider scenarios in which X-ray heating has a negligible effect on galaxy formation, as well as an example in which X-ray heating is parametrized as in \S\,\ref{sec:anal_res} with $Q_{\rm{crit}}=0.03$ and $T_{\rm{ion}}=3\times 10^4$ at $z=7$.   Table \ref{tab:Q} details $\langle Q\rangle _{\rm mass}$ for each simulation box used in this paper for which X-ray heating is ignored.  We note that $\langle Q\rangle _{\rm mass}$ remains constant at fixed epoch to within a few percent, indicating that the scheme conserves photons in the simulations described\footnote{For $\langle Q\rangle \lesssim 0.5$ fluctuations in the temperature brightness are dominated by fluctuations in the matter density field; whereas for $\langle Q\rangle \gtrsim 0.5$ fluctuations are dominated by \H2 regions.  These results, in conjunction with the output of the analytic model, have been used to interpret the output of the semi-numerical scheme.  In particular, at very high-redshift ($z\gtrsim 13$) we find that $\sigma_{21}^{2}(\theta)$ and the PS are not effected by the X-ray contribution. Similarly, at $\langle Q\rangle \lesssim 0.5$, prior to the formation of bubbles, we find that the semi-numerical scheme produces results very similar to the analytic model.}. Figure \ref{fig:Q_maps} shows slices from the three-dimensional simulation cubes at $z=7$ and $z=9$ for the fiducial (no X-ray, no heating) case and for an X-ray contribution of 30\% (no heating).  The corresponding mass-weighted ionization fractions are 0.77 and 0.32 for the fiducial case, and $0.74$ and $0.32$ for the 30\% X-ray case at $z=7$ and $z=9$ respectively.  Also represented as circles on these plots are the size of the MFP ($\approx 153$ and $98$\,cMpc at $z=7$ and $z=9$ respectively, corresponding to X-rays with mean energy 460\,eV). These maps illustrate the primary effect of adding a diffuse X-ray background. In particular the maps at $z=7$ illustrate that replacement of UV flux in the model with X-ray flux leads to smaller \H2 regions, and lower overall contrast between 21-cm emission from different regions of the IGM.  

It is important to note that the simplifying choice to consider only a single X-ray energy, rather than a polychromatic spectrum, precludes a thorough treatment of how cosmological redshift affects the spectral shape.  In particular \cite{RO04a} assert that redshifted hard X-rays constitute a large fraction of the ionizing radiation during reionization, hence affecting the topology of reionization \citep{ROG05}.  Moreover, since hard photons have a very long MFP, they can travel far from the overdensity where they are produced before redshifting to a wavelength where they are absorbed.  Thus the effect of cosmological redshift is to decorrelate the X-ray ionization with the density field even more than expected from the instantaneous X-ray SED.  These issues can only be addressed in detail by the inclusion of radiative transfer.  In lieu of radiative transfer we have assumed an effective value of MFP.  We note that a value in excess of that expected from the spectrum provides an approximation for the effects of cosmological redshift.  

The results presented in the following subsections are influenced by the interplay between three physical parameters:  (i) the MFP of X-rays, (ii) the characteristic scale of \H2 regions, and (iii) the mean ionized fraction in the simulation box $\langle Q\rangle$.  We consider first the effects of varying these three parameters without including X-ray heating.  Heating is included in the final subsection.  We briefly summarize the main effects before discussing the different cases at greater length with the aid of several examples. 

Both the MFP and the \H2 region (bubble) scale generate a shoulder in the PS, which results from the transfer of power between small and large scales.  In the case of the X-ray MFP, the transfer is across scales comparable with the MFP. Conversely, the shoulder generated by \H2 regions results from the transfer of power from small scales to scales exceeding the size of ionized bubbles.  When these two scales are comparable, it is difficult to attribute the features of the PS to a single physical process.  On the other hand, when the MFP and \H2 region scales differ substantially, the resulting PS depends on $\langle Q\rangle$.  The presence of X-rays decorrelates the density and brightness temperature, most notably on scales smaller than the MFP. Thus, as discussed in \S\,\ref{sec:anal_res}, when $\langle Q\rangle\lesssim (\gtrsim) 0.5$, the brightness temperature of overdense regions is enhanced (suppressed) relative to that of underdense regions.  In contrast, as discussed in the same section, X-ray heating of the IGM increases the bias of sources by suppressing low mass galaxies, resulting in overdense regions being ionized first, even in the presence of a large X-ray fraction.

\subsection{Models without X-ray heating}

\subsubsection{Case 1: the mean free path is smaller than the bubble scale}

\begin{figure*}
\vspace{1cm}
\includegraphics[width = 10cm,angle=90]{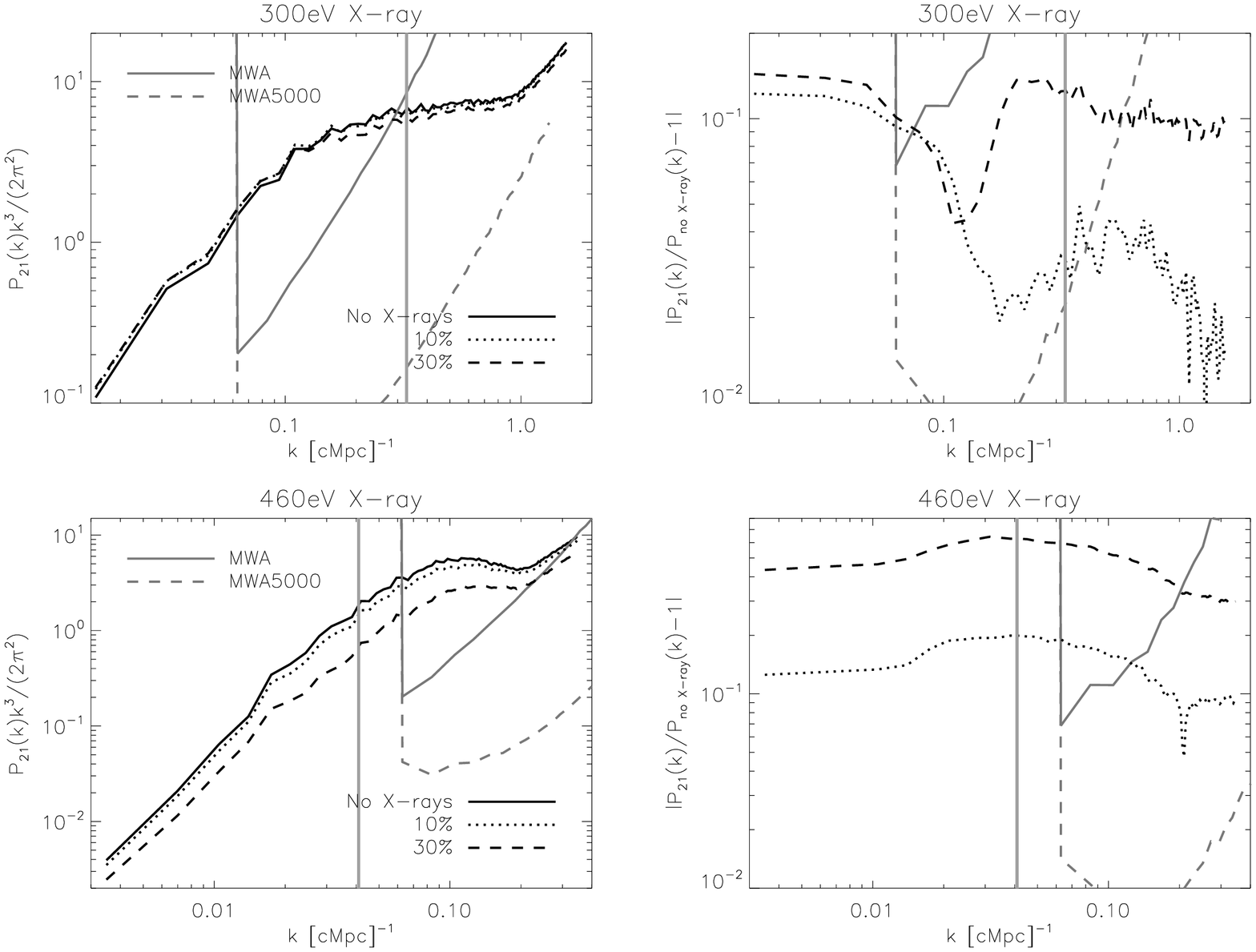}
\caption{\emph{Left}:  The power spectrum (\emph{left}) of fluctuations in the 21-cm brightness temperature at $z=7$ for X-ray contributions of 0\%, 10\% and 30\%, and mean X-ray energy of 300\,eV (\emph{top}) and 460\,eV (\emph{bottom}). The vertical line (\emph{grey}) marks the position of the MFP.  \emph{Right}:  Corresponding fluctuations relative to the fiducial no X-ray case. Also plotted for comparison is the estimated sensitivity of the MWA (upper left panels, \emph{solid grey}) and MWA5000 (upper left panels, \emph{dashed grey}), and the sensitivity relative to the PS (upper-right panels) assuming 1000\,hr integration on 1 field, and bins of width $\Delta k=k/10$. $T_{\rm{ion}}= 10^5\,\rm{K}$, $Q_{\rm{crit}} > 1$.}
\label{fig:z7_1e5}
\vspace{5mm}
\vspace{1cm}
\includegraphics[width = 10cm,angle=90]{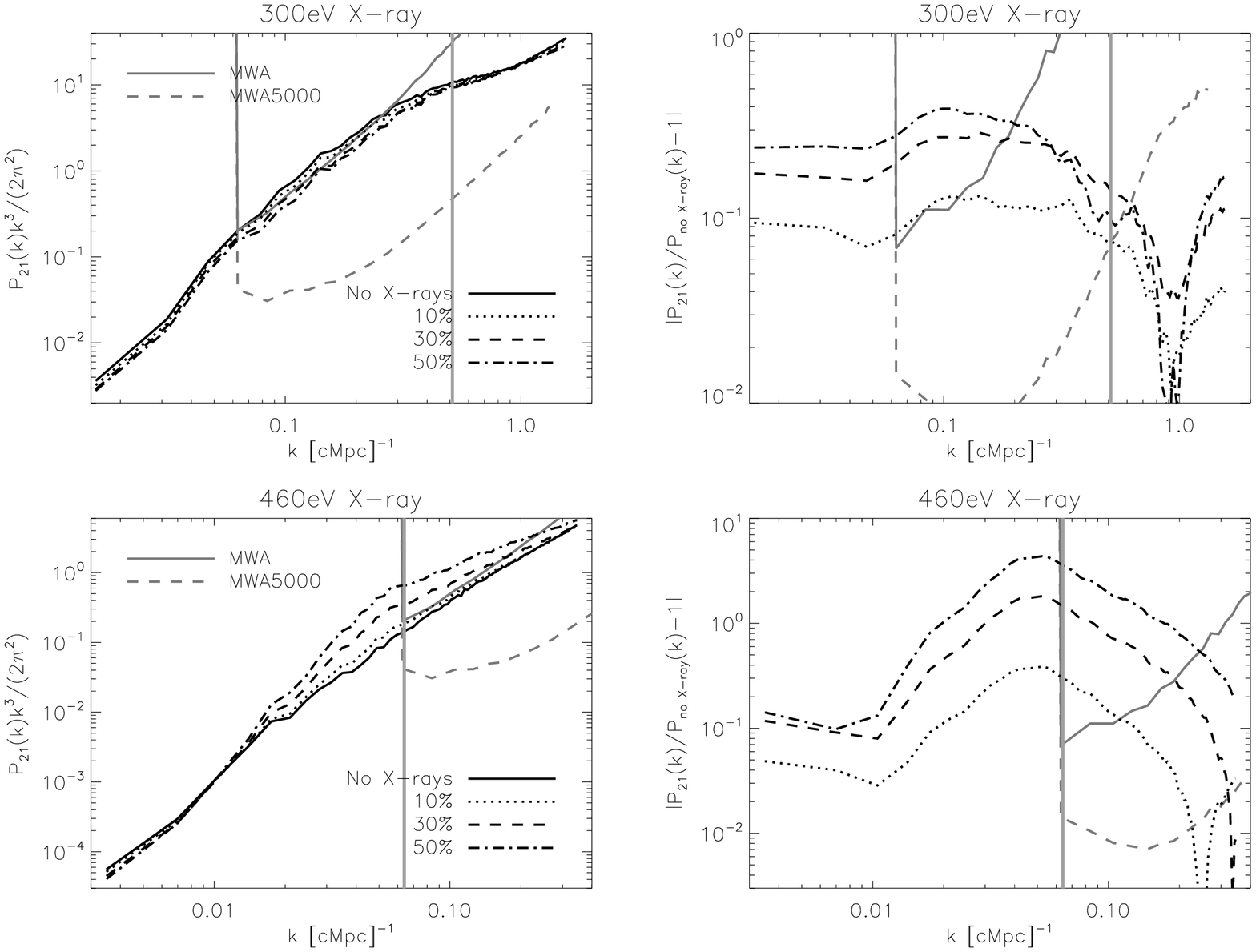}
\caption{As per Figure \ref{fig:z7_1e5} at $z=9$, with the addition of a 50\% X-ray contribution curve (\emph{dash dot}).}
\label{fig:z9_1e5}
\end{figure*}

\begin{figure*}
\vspace{1cm}
\includegraphics[width = 10cm,angle=90]{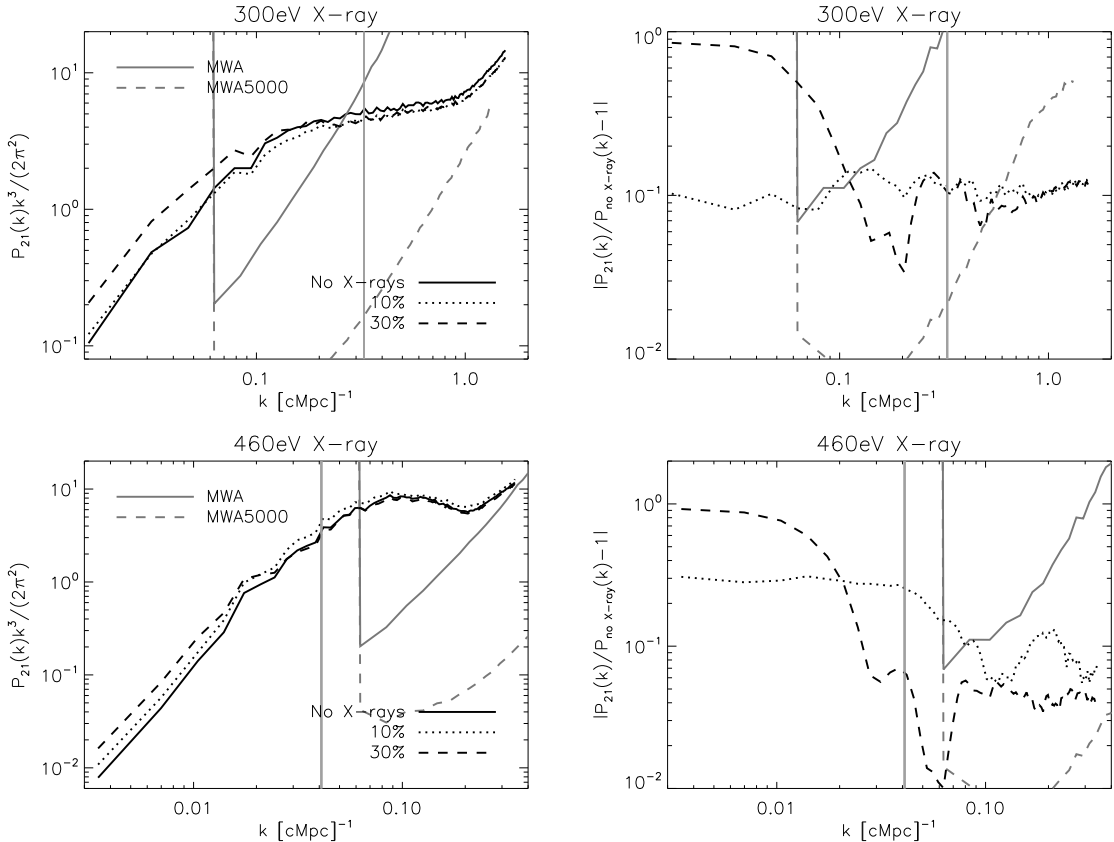}
\caption{As per Figure \ref{fig:z7_1e5} with $T_{\rm{ion}}= 3\times 10^4\,\rm{K}$, $Q_{\rm{crit}} = 0.03$.}
\label{fig:z7_3e4}
\end{figure*}

\begin{table}
\caption{Table of mean mass-weighted ionization fraction for different simulation parameters when X-ray heating is ignored.}
\begin{center}
\begin{tabular}{llllll}
\hline\hline
z & side length &$\langle E_{\rm X}\rangle_{\rm flat}$ eV & $\langle \lambda \rangle$ cMpc& $X_{\rm frac}$ & $\langle Q\rangle_{\rm mass}$ \\
\hline 
7 & 400 & 300 &	19	& 0.0 & 0.81 \\ 
7 & 400 & 300 &	19	& 0.1 & 0.80 \\ 
7 & 400 & 300 &	19	& 0.3 & 0.78 \\ \\
7 & 1800 & 460 & 153	& 0.0 & 0.70 \\ 
7 & 1800 & 460 & 153 & 0.1 & 0.70 \\ 
7 & 1800 & 460 & 153	& 0.3 & 0.69 \\ 
\hline
9 & 400 & 300  & 12	& 0.0 & 0.32 \\ 
9 & 400 & 300  & 12	& 0.1 & 0.32 \\ 
9 & 400 & 300  & 12	& 0.3 & 0.32 \\ 
9 & 400 & 300  & 12	& 0.5 & 0.31 \\ \\

9 & 1800 & 460 & 98	& 0.0 & 0.29 \\ 
9 & 1800 & 460 & 98	& 0.1 & 0.29 \\ 
9 & 1800 & 460 & 98	& 0.3 & 0.29 \\
9 & 1800 & 460 & 98	& 0.5 & 0.29 \\
\hline
\label{tab:Q}
\end{tabular}
\end{center}
\end{table}

Figure \ref{fig:z7_1e5} presents results for 21-cm fluctuations at $z=7$ for $\langle \lambda\rangle = 19\,\rm{cMpc}$ and $\langle \lambda\rangle = 153\,\rm{cMpc}$ (corresponding to $\langle E_{\rm X}\rangle = 300$\,eV and $\langle E_{\rm X}\rangle = 460$\,eV respectively for a flat spectrum).  In the upper panels of figure~\ref{fig:z7_1e5} we show the PS of 21-cm fluctuations computed from the semi-numerical model for $300$\,eV X-ray photons at $z=7$. At this late stage of reionization the MFP of $300$\,eV photons is much smaller than the \H2 bubble scale, which means that the X-ray photons that are contributing to ionizing the bubbles are being produced within the bubbles.  For this reason, the bubble scale is not affected by X-rays in this case, as illustrated by the PS shown.  However, we find that on scales smaller than the MFP, the presence of X-rays acts to suppress fluctuations in the brightness temperature. The different lines represent X-ray contributions to reionization of between 0\% and 30\% (the case of 50\% X-ray contribution is not shown because this would correspond to X-rays reionizing $\sim 40\%$ of the IGM, which exceeds the limits discussed in \S\,\ref{subsec:BH}). Also shown (\textit{upper-right panel}) is the absolute value of the fluctuation due to X-rays relative to the fiducial case (with zero X-ray contribution). 

Figure~\ref{fig:z7_1e5} demonstrates that as the X-ray fraction increases, the suppression of fluctuations below $\theta_{\langle\lambda\rangle}$ is more pronounced, in agreement with the analytic model.

\subsubsection{Case 2: the mean free path is comparable to the bubble scale}

In the upper panels of figure \ref{fig:z9_1e5} we repeat this analysis at $z=9$, when $\langle Q\rangle \lesssim 0.5$. In this case, the scale of the X-ray MFP is comparable to the characteristic bubble scale, and hence the prominence of the bubble shoulder is diminished in the presence of X-rays.  As the relative X-ray contribution increases, the bubble shoulder is less visible.  As a result of the suppression of the bubble shoulder, fluctuations on scales larger than the MFP are reduced.  Moreover, on scales smaller than the X-ray MFP, fluctuations are also suppressed, albeit to a lesser extent. Hence this result contrasts with the $z=7$ case in the upper panels of Figure \ref{fig:z7_1e5}, in which suppression of fluctuations occurred only on scales smaller than the MFP.   Once again, we invoke the decorrelation of density and ionization to explain this result.  That is, X-rays reduce the biased reionization of overdense regions by UV photons, and hence reduce the contrast between ionized regions in the IGM and the neutral component of IGM. 

The upper panels of figure \ref{fig:z9_1e5} show the case of a MFP that is comparable to the bubble scale early in reionization. However this condition is also met for a larger MFP later in reionization, as shown in the lower panels of figure~\ref{fig:z7_1e5} where the PS (\textit{left}) of 21-cm fluctuations at $z=7$ for $\langle E_{\rm X}\rangle = 460$\,eV is presented. The different lines represent X-ray contributions to reionization of between 0\% and 30\%. Also shown (\textit{right-hand panel}) is the fluctuation relative to the fiducial case. These results show that the maximum suppression of fluctuations occurs on the scale of the MFP. As the X-ray fraction increases, the shoulder due to H\,{\sevensize II} regions that is present in the fiducial PS (\emph{solid}) becomes less prominent.  As mentioned previously, this is because this feature is generated by the transfer of power from small to large scales following the formation of H\,{\sevensize II} regions, an effect which is washed out by the diffuse ionization due to X-rays.

The lower panels of figure~\ref{fig:z7_1e5} demonstrate that as the X-ray fraction increases, the suppression of fluctuations below $\langle\lambda\rangle$ is more pronounced. In accordance with the analytic results presented in \S\,\ref{sec:anal_res}, we find that the suppression of fluctuations due to X-rays is greatest at a scale slightly less than $\langle\lambda\rangle$. This is most easily seen in the plots of ratio relative to the fiducial case in the lower right panel of figure \ref{fig:z7_1e5}.

\subsubsection{Case 3: the mean free path is larger than the bubble scale}

In this subsection we explore the impact of increasing the X-ray MFP, such that the MFP is much larger than the characteristic scale of \H2 regions.  
In the lower panels of Figure \ref{fig:z9_1e5} we repeat our analysis at $z=9$, when $\langle Q\rangle \lesssim 0.5$. Most notably, the X-ray contribution magnifies fluctuations in the 21-cm signal, rather than suppressing them. As discussed in \S\,\ref{sec:anal_res}, this reversal can be attributed to the contrasting effect on the brightness temperature of underdense and overdense regions at epochs either side of $\langle Q\rangle\sim 0.5$. As at lower redshift, the scale of maximum departure from the fiducial case is slightly less than $\theta_{\langle\lambda\rangle}$.  However, unlike the low redshift case, the ratio of the power spectra asymptotes to 1 at large scales, which we attribute to there being very few ionized bubbles at high redshift. 

We find that there is the greatest qualitative agreement with the analytic results presented in figure \ref{fig:analytic_9} in this case, which we attribute to the lesser importance of \H2 regions on scales much larger than their characteristic size. Conversely, we attribute the discrepancy between the analytic and semi-numerical results closer to the end of reionization to additional fluctuations due to finite numbers of \H2 regions late in reionization, which are not captured by the analytic model.

\subsection{Models with X-ray heating:  an example}

One possible parametrization of X-ray heating is to set an ionization fraction due to X-rays ($Q_{\rm{crit}}$) above which small mass galaxy formation is suppressed.  The degree of suppression of galaxy formation is governed by the filtering scale of ionized regions ($T_{\rm{ion}}$).  Figure \ref{fig:z7_3e4} shows the results of simulations at $z=7$ for $T_{\rm{ion}}=3\times 10^4$ and $Q_{\rm{crit}}=0.03$.  As in earlier figures, we consider $300\,\rm{eV}$ (\textit{upper}) and $460\,\rm{eV}$ (\textit{lower}) X-rays, and fractional X-ray contributions of between 0\% and 30\%.  For the no-X-ray case the reduction of $T_{\rm{ion}}$ from $10^5$ to $3\times 10^4\,\rm{K}$ means that the prominence of the bubble shoulder is diminished in comparison to the case presented in the upper panels of Figure \ref{fig:z7_1e5}.  In addition, when the X-ray MFP is smaller than the bubble scale, the inclusion of X-rays increases fluctuations on all scales.  On the other hand, as discussed in \S\,\ref{sec:anal_res}, when the X-ray MFP exceeds the bubble scale (\emph{lower panels}), the profile of 21-cm fluctuations is governed by two competing effects:  an increase in the clustering bias of emission sources due to the suppression of low-mass galaxy formation; and a decrease in the bias due to the relatively long MFP of X-rays shifting ionizations from overdense to underdense regions.  Which of these two effects dominates depends on the fractional X-ray contribution to reionization. For example, when the X-ray contribution remains below $\sim20\%$, the increase in fluctuations due to the increased source bias governs the departure from the no-X-ray case.  As the X-ray contribution increases, however, fluctuations are suppressed either side of the scale of the MFP, due to the long X-ray MFP.

\section{Sensitivity of Low-Frequency Arrays to the effect of X-rays on the 21-cm PS}

\label{noise}

Before concluding, we compute the sensitivity with which the effect of
ionization by X-rays on the shape of the PS could be detected using
forthcoming low-frequency arrays.  To compute the sensitivity $\Delta
P_{21}(\mathbf{k})$ of a radio interferometer to the 21-cm PS, we
follow the procedure outlined by \cite{McQ+06} and \cite{B+06}
\citep[see also][]{WLG08}. The important issues are discussed below,
but the reader is referred to these papers for further details. The
sensitivity to the PS comprises components due to the thermal noise,
and sample variance within the finite volume of the observations. We
consider observational parameters corresponding to the design
specifications of the MWA, and of a hypothetical follow-up to the MWA
(termed the MWA5000). In particular the MWA is assumed to comprise a
phased array of 500 tiles. Each tile contains 16 cross dipoles
yielding an effective collecting area of $A_{\rm e}=16(\lambda^2/4)$
(the area is capped for $\lambda>2.1$\,m). The physical area of a tile
is $A_{\rm tile}=16\,$m$^2$.  The tiles are distributed according to a
radial antenna density of $\rho(r)\propto r^{-2}$, within a diameter
of 1.5\,km and outside of a flat density core of radius 18\,m.  The
MWA5000 is assumed to follow the basic design of the MWA. The
quantitative differences are that the telescope is assumed to have
5000 tiles within a diameter of 2\,km, with a flat density core of
80\,m. In each case we assume one field is observed for
1000\,hr. Following the work of McQuinn et al.~(2006) we assume that
foregrounds can be removed over $8$\,MHz sub-bands, within a $32$\,MHz bandpass
 [foreground removal therefore imposes a minimum on the accessible
wavenumber of $k_{\rm min}\sim0.04[(1+z)/7.5]^{-1}$Mpc$^{-1}$].

The sensitivity to the 21-cm PS per mode may
be written
\begin{equation}\label{thermal_noise}
\delta P_{21}(k,\theta) = \left[\frac{T_{\rm sky}^2}{\Delta\nu t_{\rm int}} \frac{D^2 \Delta D}{n(k_{\perp})}\left(\frac{\lambda^2}{A_e}\right)^2\right] + P_{21}(k,\theta),
\end{equation}
where $D$ is the comoving distance to the centre of the survey volume
which has a comoving depth $\Delta D$. Here $n(k_{\perp})$ is the
density of baselines which observe a wave vector with transverse
component $k_{\perp}$, and $\theta$ is the angle between the mode $\mathbf{k}$ and the line of sight. The thermal noise component (first term) is
proportional to the sky temperature, where $T_{\rm sky} \sim 250 \left(
\frac{1+z}{7}\right)^{2.6}$\,K at the frequencies of interest. The second term corresponds to sample variance. The overall sensitivity is 
\begin{equation}
\Delta P_{21}(k,\theta) = \delta P_{21}(k,\theta)/\sqrt{N_c(k,\theta)},
\end{equation}
where $N_c(k,\theta)$ denotes the number of modes observed in a $k$-space
volume $d^3k$ (only modes whose line-of-sight components fit within the observed bandpass are included). In terms of the $k$-vector components $k$ and $\theta$, $N_c = 2\pi
k^2 \sin\theta dk d\theta \mathcal{V}/(2
\pi)^3$ where $\mathcal{V} = D^2 \Delta D (\lambda^2/A_e)$ is the
observed volume. Taking the spherical average over bins of $\theta$, the sensitivity to the 21-cm PS is 
\begin{equation}
\frac{1}{[\Delta P_{21}(k)]^2}=\sum_{\theta}\frac{1}{[\Delta P_{21}(k,\theta)]^2}.
\end{equation}

The spherically averaged sensitivity curves for the MWA and MWA5000 (within bins of $\Delta k=k/10$) are
plotted as the solid grey and dashed grey lines in the upper left panels of each of Figures~\ref{fig:z7_1e5}-\ref{fig:z7_3e4}. The sensitivity as a ratio of the fiducial PS ($\Delta P_{21}/P_{21}$) is plotted in the upper right panels of these Figures (again as solid grey and dashed grey lines). 

Comparison of the sensitivity with the fluctuations induced in the PS by X-ray ionization will be measurable by the MWA in the cases of long MFP. This is because in these cases the fluctuations are largest on scales comparable to the MFP, while the low resolution of the MWA results in its sensitivity being best on large scales ($\sim0.1$\,Mpc$^{-1}$). In the examples of smaller MFPs shown, the MWA will not be sensitive to fluctuations due to X-rays, either because fluctuations are negligible at scales near $\sim0.1$\,Mpc$^{-1}$ (figure~\ref{fig:z7_1e5}) or because the fiducial PS is barely detected and the presence of X-rays suppresses, rather than enhances the fluctuation amplitude (figure~\ref{fig:z9_1e5}). In these cases, comparison with the sensitivity curve of the MWA5000 shows that detection of the fluctuations in these cases will be possible using an array with larger collecting area. 

We note that for the cases with large MFPs, the PS in this paper have been plotted out to very large scales. However, in practice the requirement that bright foregrounds be removed from the 21-cm observations prior to construction of the PS is likely to restrict observations to wave numbers $k\ga0.03-0.05$\,Mpc$^{-1}$. The lower panels of figures~\ref{fig:z7_1e5} and \ref{fig:z9_1e5} therefore illustrate that foreground removal will likely render the PS undetectable on scales comparable to and a bit larger than the X-ray MFP in these cases. As a result, the crossover from scales most effected by X-rays to scales less effected by X-rays will not be detected. However, as noted above, these figures also show that the enhancement (or suppression, depending on epoch) of the PS on scales below the X-ray MFP, which is sensitive to the fractional X-ray contribution, will still fall within observable scales. Thus we expect that measurements of the contribution of X-rays to reionization will be accessible to observations of the 21-cm PS for all MFPs, while measurements of the mean X-ray energy will only be possible for sufficiently low energies.

\section{Summary}
\label{sec:summ}

The presence of an ionizing background of X-rays imprints a scale dependent signature on the 21-cm power spectrum during reionization. The impact of X-rays on the topology of reionization depends on the redshift at which one observes, as well as the mean energy of the X-rays and their fractional contribution relative to UV emission. The findings of this investigation can be summarized as follows:
\begin{itemize}
\item  The effect of X-rays on the statistics of 21-cm fluctuations can be prominent on any scale, rather than only on scales comparable to the MFP as might be na\"{i}vely expected.
\item  In a mostly neutral IGM, X-ray ionizations enhance 21-cm fluctuations. Conversely, in a mostly ionized IGM, X-rays suppress fluctuations, provided that X-ray heating does not suppress low mass galaxy formation.
\item  ionization by X-rays modifies, and in some cases smoothes out, the signature of \H2 regions in the 21-cm PS. If reionization of the IGM included a non-negligible contribution due to X-rays, then the modification of the PS shape introduces a feature at a level that is comparable to that introduced by \H2 regions. 
\item The details of PS modification depend on the relative scales of the \H2 regions and MFP, and on the overall ionization fraction of the IGM.
\item The increase in source bias due to suppression of low mass galaxy formation by X-ray heating counterpoints the decorrelation of ionizing photons and density due to the long X-ray MFP.
\item The MWA will have sufficient sensitivity to detect the modification of the PS due to a 10-30\% contribution to reionization by X-rays, so long as the MFP falls within the range of scales over which the array is most sensitive ($\sim0.1$\,Mpc$^{-1}$). In cases where the MFP takes a much smaller value, an array with larger collecting area would be required. 
\item In some scenarios, the requirements of foreground removal will limit the observability of the effect of X-ray ionization on the PS at scales larger than the MFP. However, in all cases, the magnitude of the signature imprinted on the 21-cm PS at small scales is sensitive to the relative X-ray fraction, and will fall within the observable range. 
\end{itemize}

In conclusion, our study shows that the details of the PS shape depend on the typical X-ray photon energy, while the magnitude of any modification to the PS is determined by the relative contribution of X-rays to reionization. The possible contribution of X-rays therefore has the potential to substantially complicate analysis of the 21-cm PS. On the other hand, precision measurements and modelling of the 21-cm PS promise to provide a method for investigating the role and contribution of X-rays during reionization.

\bigskip

{\bf Acknowledgements} 
We wish to thank Steve Furlanetto for detailed comments on an earlier draft of this paper.  We also thank the anonymous referee, whose report served to improve this paper. LW and PMG acknowledge the support of Australian Postgraduate Awards. PMG is grateful for the hospitality of the Institute for Theory and Computation at the Harvard-Smithsonian Centre for Astrophysics. The research was supported by the Australian Research Council (JSBW).

\appendix

\bibliography{text_revised}
\bibliographystyle{mn2e}

\label{lastpage}
\end{document}